\documentclass[9pt,twocolumn,twoside]{opticajnl}
\journal{opticajournal} 

\setboolean{shortarticle}{false}

\usepackage{braket}

\usepackage{cleveref}
\usepackage{comment}

\newcommand{\red}[1]{{\color{black}#1}}

\title{Dynamical development of long-range spatial coherence in non-equilibrium bosonic condensation.}

\author[1,*]{Bianca Rae Fabricante}
\author[2,3,$\ddagger$]{D\k{a}br\'owka Biega\'nska}
\author[3,4] {Paolo Comaron}
\author[1]{Mateusz~Kr\'ol}
\author[1,8]{Matthias Wurdack}
\author[2]{Maciej Pieczarka}
\author[5]{Mark Steger}
\author[5]{David W. Snoke}
\author[6]{Kenneth West}
\author[3]{Daniele Sanvitto}
\author[3]{Dario Ballarini}
\author[3]{Dimitrios Trypogeorgos}
\author[6]{Loren N. Pfeiffer}
\author[1]{Andrew G. Truscott}
\author[4]{Marzena Szyma\'nska}
\author[1]{Elena A. Ostrovskaya}
\author[1,7]{Eliezer Estrecho}

\affil[1]{Department of Quantum Science and Technology, Research School of Physics, The Australian National University, Canberra ACT 2601 Australia}
\affil[2]{Department of Experimental Physics, Faculty of Fundamental Problems of Technology, Wrocław University of Science and Technology, Wyb. Wyspianskiego 27, 50-370 Wrocław, Poland }
\affil[3]{CNR Nanotec, Institute of Nanotechnology, via Monteroni, 73100, Lecce, Italy}
\affil[4]{Department of Physics and Astronomy, University College London, London, United Kingdom}
\affil[5]{Department of Physics and Astronomy, University of Pittsburgh, Pittsburgh, Pennsylvania 15260, USA}
\affil[6]{Department of Electrical Engineering, Princeton University, Princeton, New Jersey 08544, USA }
\affil[7]{School of Science, The University of New South Wales, Canberra ACT 2612, Australia}
\affil[8] {Department of Chemical Engineering, Stanford University, Stanford, CA, USA}

\affil[$\ddagger$]{Present address: Turner Institute for Brain and Mental Health, School of Psychological Sciences, and Monash Biomedical Imaging, Monash University, Clayton, Victoria, Australia}
\affil[*]{bianca.fabricante@anu.edu.au}

\begin{abstract}
Development of spontaneous coherence is one of the hallmarks of bosonic condensation in a variety of physical systems, such as cold atoms, confined photons, and hybrid light-matter quasiparticles like exciton polaritons in semiconductors. 
While spatial coherence is well understood once a steady-state condensate has been established, its temporal evolution as the condensate forms is largely unexplored. 
Here, we explore the dynamical formation of a non-equilibrium, driven-dissipative exciton-polariton condensate through both time-resolved experiments and numerical modeling. Our study reveals that the spatial coherence is established through two distinct stages. 
The early-time stage is interaction-driven and features transient oscillations in spatial coherence. 
This stage is followed by a steady-state regime characterized by a spatially-uniform high degree of coherence that extends over the entire size of the system and persists over time. 
These stages of spatial coherence development occur in both free flowing and confined exciton-polariton systems that undergo a quench --- rapid growth of the condensate starting from two different initial settings. 
Our study offers a deep insight into the process by which long-range spatial coherence is established in a non-equilibrium bosonic condensate. 
\end{abstract}

\setboolean{displaycopyright}{false}

\begin{document}
\doi{}
\maketitle

\section{Introduction}

The emergence of order from disorder is one of the most fundamental processes in nature. The questions of how the coherence builds up and how the system selects its ordered state transcend the scales of physical processes from cosmological symmetry breaking during inflation \cite{linde1982,kibble1976,zurek1985cosmological} to quantum phase transitions in table-top experiments \cite{weiler2008spontaneous}. These vastly different dynamical processes are unified by the same underlying framework of spontaneous symmetry breaking and critical dynamics \cite{HH1977,Bray01061994}. A particularly powerful way to probe these dynamics is through a \textit{quench} — a sudden drive across the phase transition that exposes how fluctuations seed the emerging order, how coherent domains nucleate and coarsen, and how long-range coherence gradually develops in an initially disordered state~\cite{Bray01061994}. In the past few decades, bosonic condensates of atoms, photons, and light-matter excitations in semiconductors (exciton polaritons) have provided an accessible experimental platform for exploring these universal questions.

In cold-atom bosonic condensates, where the system reaches thermal equilibrium, coherence formation is well understood, and the development of spatial coherence directly reflects the build-up of phase correlations across the system~\cite{Navon167,chomaz2015}. In driven-dissipative systems, where condensation is sustained by a balance of gain and loss rather than thermal equilibrium, these questions become especially complex and remain largely unexplored experimentally.

Exciton polaritons (polaritons herein) are hybrid light-matter particles that can be formed by the strong coupling between confined excitons (electron-hole pairs) in direct bandgap semiconductors and photons confined, e.g., in a microcavity. Polaritons inherit the light effective mass and extended coherence times of photons and interact via Coulomb repulsion of constituent excitons. Their short, picosecond-scale lifetime means that injection of excitons (via laser excitation) is needed in order to reach and maintain the density sufficient for condensation. On the other hand, their radiative decay enables the study of the condensates {\em in-situ}, by capturing and analyzing the resulting photoluminescence. These unique attributes make the system a rich test bed for probing non-equilibrium bosonic condensation~\cite{kasprzak2006bose, deng2010exciton, Balili1007, matuszewski2014universality, Snoke2012}, development of spontaneous coherence~\cite{Deng2007, richard2005spontaneous,Kim2016,Laussy2004,Comaron2024,Demenev2016,Roumpos2012PNAS,Klaas2018}, superfluidity~\cite{Kolmakov:16, keeling2016superfluidity, amo2009superfluidityCompletion, Rubo2006, Lagoudakis2008, amo2009superfluidityFirst}, and non-Hermitian physics~\cite{Liang2026, Xu2025, RobinHu2024, RobinHu2025, Su_ScienceAdvances2021, Gao_ControlledOrdering_2018, Gao2015NonHermitian2015}. 
This system also offers a plethora of potential applications in low-threshold lasing~\cite{Zhang2022,deng2003polariton,schneider2013electrically, bhattacharya2014room, Rui2017, kenacohen2010, BiancaF2024}, quantum photonics~\cite{Delteil_2019, Kang2023}, quantum computing~\cite{kavokin2022polariton, Ghosh2020, barrat_qubit_2023, Kyriienko2016}, and topological photonics ~\cite{Beierlein2024, Jin2024, Whittaker2021}. 

Spontaneous long-range spatial coherence is at the heart of most applications based on polaritons~\cite{Kim2016, barrat_qubit_2023, Zhang2022, deng2003polariton, schneider2013electrically, bhattacharya2014room, Rui2017, kenacohen2010}. It develops as a consequence of the inherently non-equilibrium, driven-dissipative process of condensation, which makes it a non-trivial phenomenon. The spatial coherence of a polariton system not only serves as an evidence of bosonic condensation, but also offers important insights into the interplay of the system's vital parameters that often cannot be measured directly. Two of such key parameters are the density of the incoherent excitonic reservoir, which is injected by an optical pump and drives the condensation process, and the rate of energy relaxation (thermalization) in the system \cite{kasprzak2006bose, Deng2006}. Both the emergence of the polariton condensate and the growth of its spatial coherence are heavily influenced by these parameters.

\begin{figure}[t!]
    \centering
\includegraphics[width=\columnwidth]
    {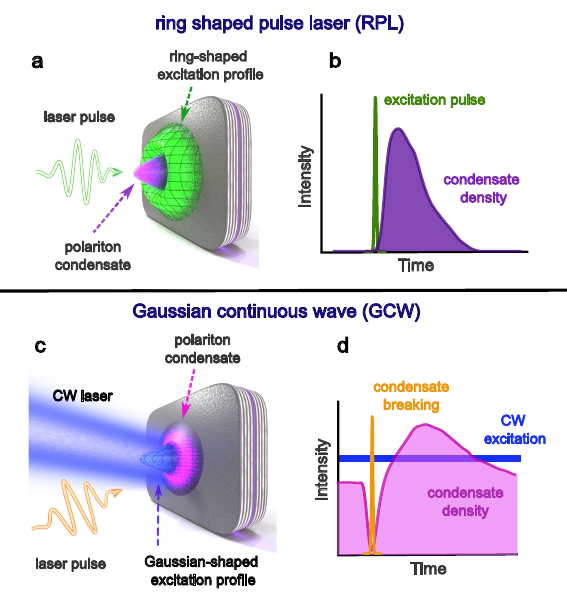}
    \caption{Two experimental configurations. a) In the ring-shaped pulsed laser (RPL) pump configuration, a pump pulse creates the dynamical polariton condensate trapped inside a ring-shaped excitonic reservoir generated by the pump. The condensate forms after the arrival of the pulse and subsequently decays, as shown in (b). c) In the Gaussian continuous wave pump (GCW) configuration, a steady-state condensate flowing away from the excitation spot is formed. A red-detuned laser pulse is then used to break the condensate. The pulse initiates a rapid polariton density depletion followed by a rapid growth and recovery of the condensate, as shown in (d).}

    \label{fig:1}
\end{figure}

In two-dimensional (2D) bosonic systems, a quasi-long range order forms across the condensate upon crossing a threshold density, following the Berezenskii-Kosterlitz-Thouless (BKT) mechanism \cite{Roumpos2012}, which is characterized by an algebraic decay of spatial coherence \cite{nitsche2014algebraic, caputo2018}. 
However, in finite-sized 2D polariton systems the coherence decay follows a stretched exponential law \cite{Comaron2021} due to the non-equilibrium nature of polariton condensates.
The formation of spontaneous coherence in a polariton condensate can be viewed as a dynamical process whereby different regions of the condensate become phase correlated. The crucial early stages of this phase ordering depend on microscopic processes governing polariton energy relaxation and interactions, which play a key role in polariton condensation under the non-resonant optical excitation. In this regime, a pump laser is used to inject excitons into the system, which undergo energy relaxation, couple with photons, and form high-energy polaritons with a high excitonic fraction. These accumulate in an incoherent reservoir until the critical density is reached and triggers the interaction-driven stimulated bosonic scattering into the lowest energy state, where the coherent condensate forms \cite{carusotto2013quantum,   kasprzak2006bose, nitsche2014algebraic, Brune2025QuantumCoherence, Nardin2009, MilenaDGiorgi2014, BiancaF2024}. 

Here, we gain insight into the formation and evolution of spatial coherence as the polaritons undergo rapid growth of the condensate (a quench) under non-resonant excitation. We observe coherence oscillations during the early stages after the quench that can be either suppressed or enhanced by tuning basic condensation parameters that control the strength of polariton interactions. Our extensive numerical modeling suggests that these oscillations arise from the dynamical redistribution of the polariton population between the ground and excited energy states, which is a unique feature of the condensate formation in this non-equilibrium system. Once the oscillations disappear, a steady and uniform build-up of spatial coherence follows. By performing the coherence measurements and respective simulations in two different experimental configurations, we confirm that the observed behavior  occurs in both free-flowing ~\cite{Deng2007,PhysRevLett.118.215301} and optically confined ~\cite{orfanakis_ultralong_2021, BiancaF2024, Askitopoulos2019} condensates, and does not depend on the state of the polaritons before the quench.

\section{Experimental details}

The samples used in both experimental configurations are high quality (high Q-factor) planar microcavities formed by distributed Bragg reflector mirrors with embedded exciton-hosting GaAs quantum wells~\cite{Steger_polaritonlifetime} (see Supplemental Document \cite{SM}).

In this work, we explore two experimental configurations, schematically presented in Fig.~\ref{fig:1}, with the respective experiments performed at two different research laboratories. The first is the ring-shaped pulsed laser (RPL) pump configuration that uses a pulsed, off-resonant laser excitation to generate the condensate as sketched in Fig.~\ref{fig:1}a-b. The second is the Gaussian continuous-wave (GCW) pump configuration, where a steady-state condensate generated by an off-resonant continuous-wave (CW) Gaussian beam is briefly disrupted by a pulsed laser and then forms again under the CW excitation Fig.~\ref{fig:1}c-d.

In the RPL configuration, the sample is pumped off-resonantly with a femtosecond pulsed laser (blue-detuned by $\sim 100$~meV from the polariton resonance) to ensure that any coherence from the pump is lost through a complex energy relaxation process~\cite{byrnes2014exciton}. 
The pump beam is shaped into a ring and then imaged onto the sample~\cite{estrecho2019direct}. 
This creates a ring-shaped incoherent excitonic reservoir that repels polaritons towards the center of the ring, where they form a condensate~\cite{askitopoulos2013polariton}, which is effectively trapped and separated from the bulk of the reservoir~\cite{askitopoulos2013polariton,estrecho2019direct,sun2017direct}, as shown in Fig. \ref{fig:1}a. Both the reservoir and the condensate decay within $1$ ns, before the next pulse arrives $12.5$~ns later, with the polaritons emitting photons that escape the microcavity as photoluminescence (PL). Therefore, within a single condensate realization, neither the reservoir nor the condensate are replenished by the pump before they decay. This configuration represents a quench from a 'quasi-vacuum' state, i.e. a rapid transition to condensation starting from a state initially populated by thermal noise, see Fig. \ref{fig:1}b.

In the GCW configuration, we use a non-resonant Gaussian CW laser beam focused on the sample to create an expanding, free-flowing condensate on top of the Gaussian-shaped reservoir ~\cite{PhysRevLett.118.215301}, see Fig. \ref{fig:1}c. We then use a femtosecond pulsed laser, red-detuned from the excitonic resonance and superimposed in space with the CW excitation, to break the condensate, see Fig.~\ref{fig:1}d. Acting via an optical AC Stark effect, the pulse strongly shifts the excitonic line, effectively temporarily removing the excitonic reservoir, leading to an abrupt loss of the strong coupling and the subsequent density depletion of the condensate \cite{Hayat2012,Forero2021}. As the CW excitation continuously pumps the system, the condensate builds up again and returns to a steady-state density after the pulse. Due to the density depletion of the condensate induced by the pulse, this configuration also represents a quench from a quasi-vacuum state, but the transition to condensation occurs under CW rather than pulsed excitation.

Time-resolved measurements of the condensate distribution, energy, and spatial coherence are performed on the condensate PL using a streak camera in conjunction with a spectrometer and a Michelson interferometer.

\begin{figure}[t!]
    \centering
\includegraphics[width=\columnwidth]
{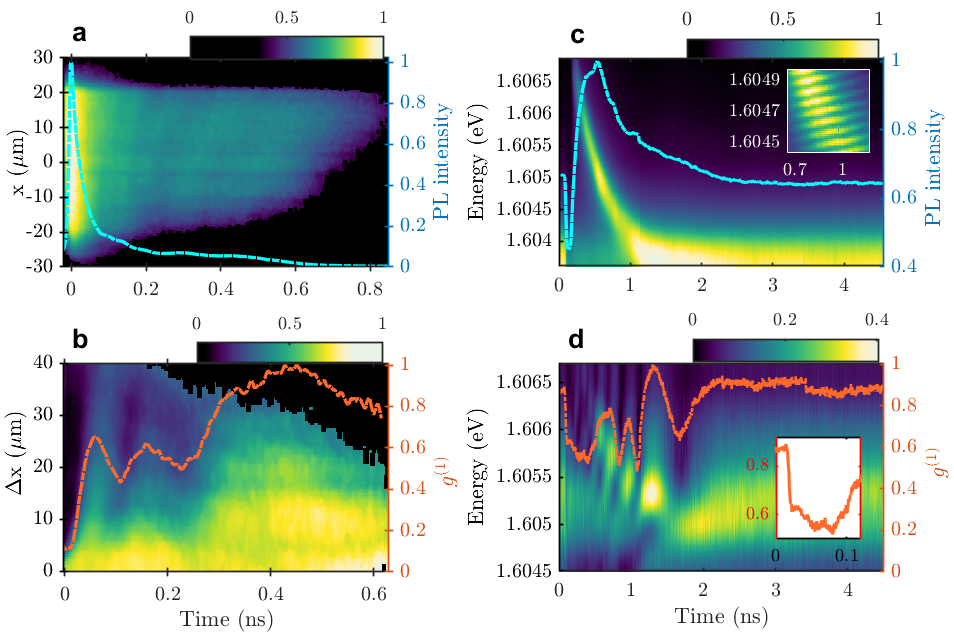}
    \caption{Typical time evolution of the condensate density and $g^{(1)}$ for (a,b) RPL and (c,d) GCW configurations, with the ratios of the pump power relative to the condensation threshold $P/P_{th}=26$ and $B_{p}/B_{p,0}=12$, respectively. a) Condensate density evolution (quantified by PL intensity heatmap) and its value integrated over the spatial ($x$) dimension and normalized by the peak value (blue line). b) Condensate coherence heatmap (shown for half of the spatial range, $x>0$) and $g^{(1)}$ value integrated over the spatial coordinate ($x$) and normalized to its peak value (orange line). c) Condensate density evolution, quantified by the time-resolved PL intensity, and the total detected PL intensity at each time, normalized to its peak value (blue line). d) Condensate coherence heatmap and the $g^{(1)}$ value integrated along the vertical axis and normalized to its peak value after the quench (orange line). The drop in spatial coherence induced by the condensate-breaking pulse is shown in the inset.}
    \label{fig:2}
\end{figure}

 In the RPL configuration, the measurements unavoidably average over thousands up to millions of pulses to increase the signal-to-noise ratio, with the averaging washing out any stochastic dynamics. Example of the time evolution of the condensate spatial distribution in the RPL configuration is presented in Fig. \ref{fig:2}a. Here, $t=0$ corresponds to the peak of the condensate PL, and the PL intensity is proportional to the condensate density.

In addition, we employ Michelson interferometry to measure the $1^{st}$-order correlation function $g^{(1)}$, which quantifies the spatial coherence of the condensate~\cite{caputo2018}. 
The condensate emission is split into two beams $I_1(\mathbf{r},t)$ and $I_2(\mathbf{r},t)$ using a beam splitter, with one beam retro-reflected. 

The beams are then recombined in such a way that they form vertical interference fringes, and are  guided towards the horizontal slit of the streak camera for time-resolved measurements. 
The horizontal $g^{(1)}$ profile cutting through the center of the condensate is given by
\begin{equation}
    g^{(1)}(\Delta x, t,\Delta t =0) = \frac{\langle \Psi(x,t)   \Psi^\dagger(-x,t) \rangle}{\sqrt{\langle |\Psi(x,t)|^2 \rangle \langle  \Psi^\dagger(-x,t)|^2 \rangle}} ,
\label{eq:g1}
\end{equation}
where $\Psi$ is the condensate wavefunction.
Here, $\Delta x$ is the distance between two points of the condensate, measured from the center, i.e. $\Delta x = 2x$. 
Note that the time delay $\Delta t$ between the two beams is kept near zero, so the measured $g^{(1)}$ represents the spatial coherence as a function of time $t$. Using the $g^{(1)}$ measurement, we can additionally quantify the coherence length, $\ell_{\rm coh}$, according to the formula:
\begin{equation}
   g^{(1)}(\Delta x,t) \sim e^{-\frac{\Delta x}{\ell_{\rm coh}}}.
   \label{eq:cohlength}
\end{equation}
(see Supplemental Document \cite{SM}).

An example of the spatial coherence measured across one of the spatial coordinates ($x$) in the RPL configuration is shown in Fig.~\ref{fig:2}b. Note that at later times, (t $\geq$ 0.6~ns), the condensate emission becomes too weak for a reliable extraction of $g^{(1)}$. The condensate also shrinks as it decays (see Fig.~\ref{fig:2}a), resulting in the narrowing of the $g^{(1)}(x)$ profiles at later times.

In the GCW configuration, in order to increase the visibility of the interference fringes and to filter out the remaining light from excitation lasers, we spectrally dispersed the signal with the use of the monochromator. However, because the slit is opened to collect emission from the entire condensate, its finite extent limits the effective spectral resolution to approximately 0.5~meV (see Supplemental Document \cite{SM}). Fig.~\ref{fig:2}c shows the measured time-resolved interferogram from which the integrated PL intensity (quantifying the condensate density) and $g^{(1)}$ (Fig.~\ref{fig:2}d) are extracted. The vertical axis of the PL intensity (Fig.~\ref{fig:2}c) and spatial coherence (Fig.~\ref{fig:2}d) plots has details of both space and energy. With a constant energy of the single-mode condensate, the vertical axis can be regarded purely as the spatial dimension, however, any change of the condensate energy can be observed as a global shift of the interference pattern on the streak camera’s slit, due to the diffraction grating. Similar to the RPL scheme, the measurements are averaged over many experimental realizations of the condensation process following a condensate-breaking pulse.

\section{Coherence dynamics}

The time evolution of the spatial coherence for the two experimental configurations presented in Fig.~\ref{fig:2}b,d occurs on different time scales. In the RPL configuration, the spatial coherence dynamics is fast and ends at $\sim$0.7~ns since the condensate decays without a constant pump. The GCW configuration allows for continuous replenishment of the condensate resulting in a long-lived coherence. In order to follow the long-term evolution of the condensate density (PL intensity) and coherence in this regime, we measure the temporal dependence of the interference fringes at later delays after the pulse, which allows us to follow the dynamics up to several nanoseconds in this configuration. Surprisingly, despite the difference in timescales, the coherence dynamics exhibit a consistent behavior, which displays much richer dynamics than the monotonic rise and decay of the condensate density shown in Fig.~\ref{fig:2}a,c. In the following, we describe this behavior in different configurations.

\subsection{Two stages of coherence development}

In the RPL configuration, after the arrival of a pump pulse, a condensate forms at the center of the ring-shaped trap formed by the reservoir and rapidly reaches a large density that manifests in a large energy blueshift~\cite{deng2010exciton, estrecho2019direct}. 
Both the reservoir and the condensate subsequently decay, with the reservoir continuing to feed the condensate while decaying over $\sim 1$ ns. This results in the condensate lifetime massively exceeding the polariton lifetime ($\sim 100$~ps) in this sample. As the condensate decays, its density, $|\psi(x,t)|^2$, decreases monotonically (as shown in Fig.~\ref{fig:2}a) and its energy redshifts. Full details on the formation and decay dynamics of the polariton condensate are provided in Supplemental Document \cite{SM}.

In contrast with the monotonic condensate density decay, the time evolution of coherence in the RPL configuration involves two regions with distinct behavior, as shown in Fig.~\ref{fig:2}b. The first one features the growth, partial collapse, and revival of spatial coherence, hereafter referred to as transient coherence oscillations (TCO). This regime occurs at early times, and is followed by the quasi-steady state (QSS) regime (at $t\gtrapprox0.3$~ns) where coherence grows to its maximum value and extends to the full size of the condensate. Subsequent slow decay of the coherence is due to the condensate decay in this regime.
In the GCW configuration, the pulse both depletes the condensate (see Fig.~\ref{fig:2}c) and destroys the associated coherence (see inset in Fig.~\ref{fig:2}d). After the pulse, and similarly to the RPL configuration, the condensate rapidly reaches a high-density, strongly blue-shifted state under the non-resonant CW excitation. It then flows away from the excitation region, being repelled by the Gaussian potential barrier created by the reservoir, and its density decays monotonically to its initial non-zero steady-state value defined by the CW pump power (see Fig.~\ref{fig:2}c). In contrast with this monotonic behavior, and similarly to the RPL configuration, the evolution of spatial coherence features a TCO regime (up to $\sim 2$ ~ns) followed by a QSS regime, where the spatial coherence stabilizes to a nearly constant value.

In line with the different time scales of the condensate evolution, the transition between the TCO and QSS regimes occurs at different timescales for the two configurations; $t\lessapprox$0.3~ns for RPL and $t\lessapprox$3~ns for GCW. In the RPL configuration, as seen in Fig.~\ref{fig:2}b, the first rise and fall of coherence occurs within $t$$\approx$~0.1~ns, followed by an abrupt (faster) loss within 30~ps. 
This is followed by another albeit weak rise and fall of coherence. At $t$~$>$~0.2~ns, the coherence finally rises again towards the QSS regime. Similarly, in the GCW configuration, the TCO regime features several oscillations in the $g^{(1)}$ temporal dependence: clear minima can be seen at around $t=0.4, 1.0, 1.3, 2.2$~ns after the pulse and the QSS regime extends up to the experimental measurement limit ($\sim 4$~ns), see Fig.~\ref{fig:2}d.

\subsection{Long-time coherence dynamics}

The QSS regime in the RPL configuration features a relatively flat spatial coherence profile shown in Fig.~\ref{fig:2}b. The very slow decay of spatial coherence with $x$ strongly indicates that the coherence extends throughout the condensate, only limited by the condensate's finite size. By fitting Eq.~(\ref{eq:cohlength}) to the $g^{(1)}$ at $t~=~504$~ps, we extract the coherence length of $\ell_{\rm coh}=921~\mu$m, which is more than 20 times wider than the condensate diameter (see Fig. S8 in Supplementary Document \cite{SM}). Previous works also reported this slow decay of $g^{(1)}$ in dynamical polariton condensates, wherein coherence reaches its maximum value shortly after the condensate population peaks, and outlasts the condensate's population decay \cite{Nardin2009,mylnikov2015}. In our case, however, spatial coherence peaks and flattens at a later stage, when the condensate density has decayed to $\sim 10 \%$ of its original value.

The phenomenology of the TCO to QSS transition is very similar for the GCW configuration~\cite{Comaron2024,nitsche2014algebraic,Roumpos2012PNAS} despite the difference in the timescales. Interestingly, during the TCO regime, the $g^{(1)}$ value increases above the initial level and becomes the highest at $\sim 1.7$\,ns after the pulse. Once the system enters the QSS regime, $g^{(1)}$ stabilizes at the steady-state value achieved before the arrival of the condensate-breaking pulse.

We stress that, in both configurations, unlike the spatial coherence, the real-space density profile does not feature oscillations, as seen in Fig.~\ref{fig:2}a,c. The absence of density oscillations indicates that the observed early-time TCO dynamics and TCO-QSS transition at a later time is a consequence of an underlying dynamical process not captured by the measurements that effectively average over multiple realizations of the condensate.

To the best of our knowledge, such dynamical coherence behavior in the transient and steady-state regimes have not been observed before.

\begin{figure*}[t!]
    \centering
    \includegraphics[width=\textwidth]{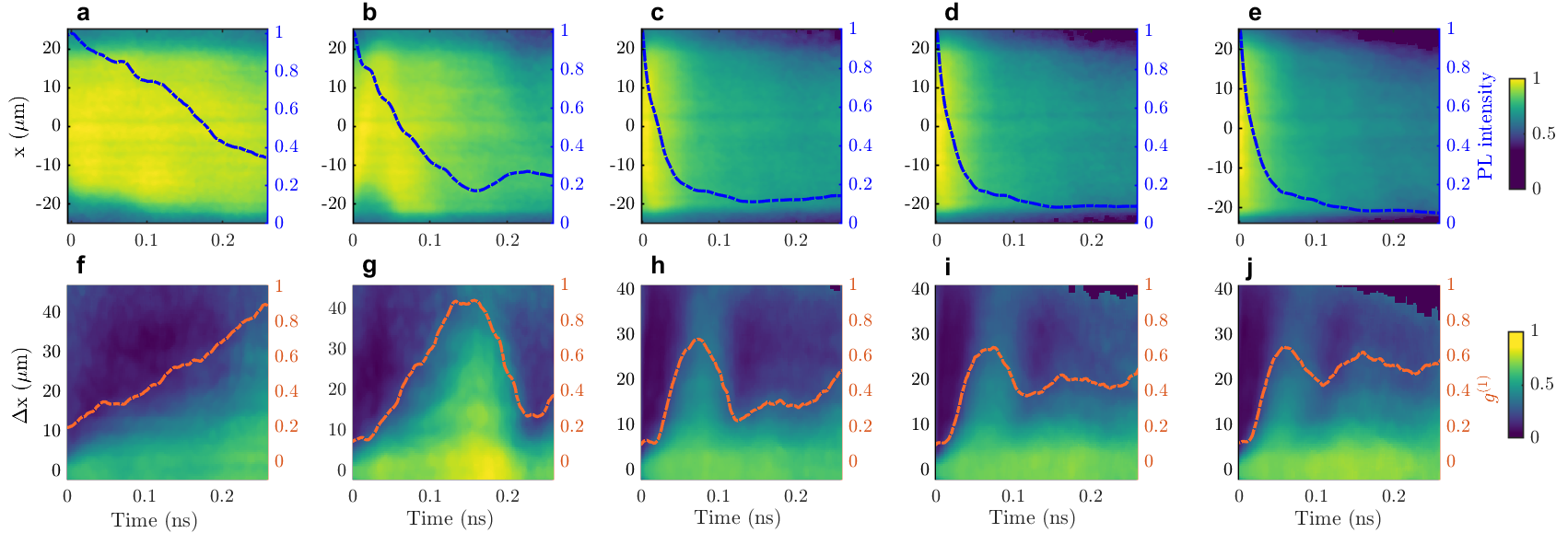}
    \caption{{Time-resolved condensate PL and $g^{(1)}$ heatmaps for the excitonic fraction $|X|^2~=~0.50$ and the pump power ratio $P/P_{th}$ of (a,f) 4, (b,g) 6, (c,h) 17, (d,i) 22, and (e,j) 26 in the TCO regime of the RPL experiment. The blue line is the condensate density quantified by integrating the PL intensity over the spatial ($x$) dimension and normalizing it by the peak value. The orange line is the spatial coherence integrated over the spatial coordinate ($x$) and normalized to its peak value.}}
\label{fig:3}
\end{figure*}

\begin{figure*}[t!]
    \centering
    \includegraphics[width=\textwidth]{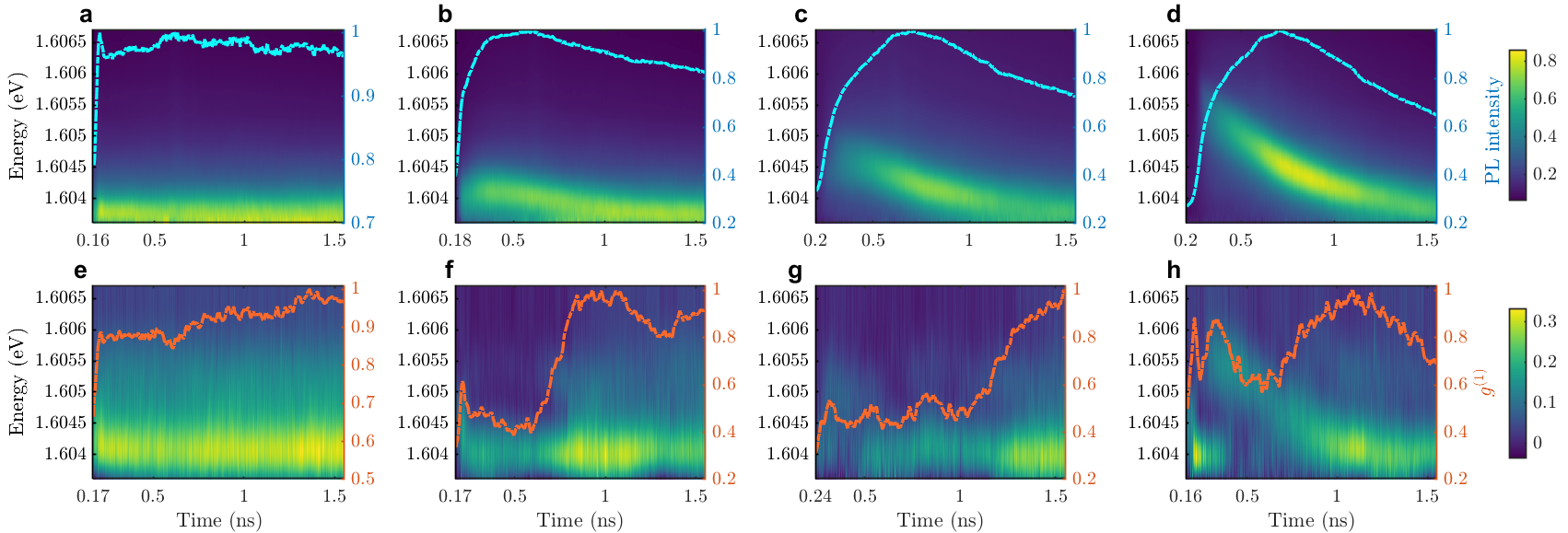}
    \caption{{Time-resolved condensate PL and $g^{(1)}$ heatmaps for the breaking pulse power ratio $B_{p}$/$B_{p,0}$ of (a,e) 1, (b,f) 2, (c,g) 4, and (d,h) 8 in the TCO regime of the GCW experiment, after the condensate-breaking pulse. $B_{p,0}$ is the threshold pulse power for breaking the condensate. The blue line is the total detected PL intensity at each time normalized to its peak value. The orange line is the spatial coherence integrated along the vertical axis normalized to its peak value.}}
\label{fig:4}
\end{figure*}

\section{Robustness of dynamical behavior}

To gain more insight into the observed TCO regime, we tuned various experimental parameters, namely the pulsed pump power,$P$, in the RPL configuration, the breaking pulse power, $B_{p}$, in the GCW configuration, and the excitonic fraction of the polaritons, which controls the strength of interparticle interactions (see Supplemental Document \cite{SM}). The pulse power in both configurations effectively controls the time scale of the condensation transition, thus determining the strength of the quench.  

\subsection{Strength of the quench}

\subsubsection{Pulsed pump power in RPL configuration}

The power of the RPL pump pulses determines the initial density of the excitonic reservoir that confines and feeds the condensate, which consequently determines the peak density of the condensate (and interaction energy) and the timescales of condensation, such as the growth and decay of coherence. The early-time behavior of the coherence and density dynamics as a function of pump power relative to the condensation threshold, $P/P_{th}$, is presented in Fig.~\ref{fig:3}a-e for a constant excitonic fraction of the polariton characterized by the excitonic Hopfield coefficient $|X|^2~=~0.50$ (see Supplemental Document \cite{SM}).

At low pump powers $P<4$~$P_{th}$ (see Fig.~\ref{fig:3}a), the TCO regime is absent. Spatial  coherence initially grows slowly up to  $t\sim0.2$~ns, after which it grows rapidly to cover the whole condensate and reach the QSS regime. 
This is true for all excitonic fractions explored in the experiment (see Supplemental Document \cite{SM} for more details). The slow buildup of coherence is expected for pump powers close to the condensation threshold, $P_{th}$, since the energy relaxation of polaritons towards the ground state is slower and less efficient resulting in a slower build-up of spatial coherence~\cite{Belykh2013}.

At higher pump powers $P>4~P_{th}$, the abrupt loss of spatial coherence starts occurring (see Fig.~\ref{fig:3}b), which is the characteristic behavior in the TCO regime discussed above. With increasing pump power ($P>10~P_{th}$), the TCO regime occurs earlier and includes multiple rise and decay periods as can be seen in Fig.~\ref{fig:3}c-e. Interestingly, the time of the TCO-QSS transition remains largely unchanged (see Fig.~S6  of \cite{SM}).

These results suggest that the increase in excitation pump power drastically changes the TCO regime but has no effect on the QSS regime.

\subsubsection{Condensate-breaking pulse power in GCW configuration}

Unlike the excitation pulse in the RPL configuration, the condensate-breaking pulse in the GCW configuration acts on a condensate continuously sustained by the CW pump. Fig. ~\ref{fig:4} shows the time-resolved PL and $g^{(1)}$ following the arrival of the breaking pulse for increasing breaking pulse power $B_{p}/B_{p,0}$, where $B_{p,0}$ is the threshold power of the condensate-breaking pulse. The blue curves show the total detected PL intensity at each time, normalized to its peak value, while the orange curves show the corresponding integrated $g^{(1)}$, also normalized to its peak value. The full dynamics, including the steady-state before the quench, are shown in the Supplemental Document \cite{SM}. At $B_{p}/B_{p,0}$ = 1 (Fig.~\ref{fig:4}a,e), the breaking pulse produces a rapid depletion of the condensate PL together with a strong reduction of spatial coherence.  Above this threshold, increasing the pulse power has only a weak effect on the magnitude of the initial condensate depletion. The condensate subsequently recovers under the continuous CW excitation, while the coherence dynamics becomes increasingly sensitive to the breaking-pulse power.

At higher pulse powers, the revived condensate exhibits a transient energy blueshift visible as the displacement of the PL and interference signal in Fig.~\ref{fig:4}b-d. The magnitude of the blueshift increases with pulse power and is accompanied by an enhancement of the PL intensity following the pulse. This behavior is most likely associated with the creation of an additional excitonic reservoir through the high-energy tail of the pulsed laser and/or two-photon absorption processes \cite{Hayat2012, Forero2021}.

The increasing breaking-pulse power has an even more pronounced effect on the coherence dynamics. At lowest power, $g^{(1)}$ recovers towards its initial value without pronounced oscillations. With increasing $B_{p}/B_{p,0}$, transient oscillations emerge and become progressively more pronounced (Fig.~\ref{fig:4}f-h). Thus, while the initial condensate depletion changes little once the breaking threshold is reached, the subsequent energy, density, and coherence dynamics depend strongly on the pulse power.

\subsection{Excitonic fraction}

The excitonic fraction affects the polariton interaction strength~\cite{Sun2017, estrecho2019direct}, energy relaxation~\cite{estrecho2018single,Comaron2024, deng2010exciton}, and effective mass. As a result, all of these parameters are effectively controlled in our samples through varying exciton-photon detuning $\delta = E_c-E_{ex}$ (see Supplemental Document \cite{SM}). 
Polaritons with a higher excitonic component (more positive detuning) relax more efficiently to the ground state~\cite{deng2010exciton, estrecho2019direct} and the condensation threshold is reached at lower densities~\cite{deng2010exciton}. 
These changes in the condensation dynamics affect the way spatial coherence is established within the condensate, providing more insight into the observed dynamical behaviour.

\begin{figure}[t!]
    \centering
    \includegraphics[width=\columnwidth]{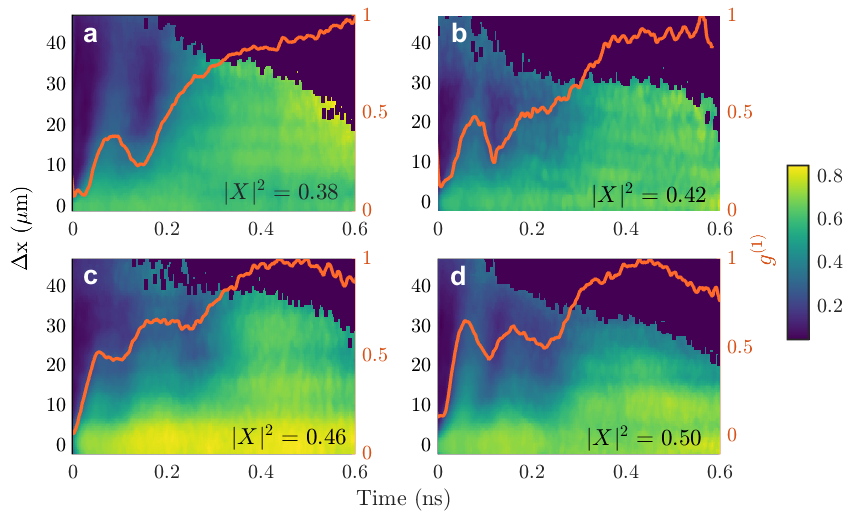}
    \caption{Dependence of spatial coherence dynamics on the excitonic fraction in the RPL configuration. (a-d) Time-resolved $g^{(1)}$ profile heatmaps at increasing (from a to d) excitonic fraction with a fixed pump power ($P/P_{th}\approx 26$). The orange curve is $g^{(1)}$ integrated (along $x$) and normalized to its peak value.}
    \label{fig:5}
\end{figure}

To observe the effect of increasing excitonic fraction on the spatial coherence dynamics, we used the RPL configuration. We varied the exciton-photon detuning from $-3.9$~meV (excitonic Hopfield coefficient ${|X|}^2 = 0.38$) to zero detuning (${|X|}^2 = 0.50$) and kept the pump power at  $P/P_{th}\approx 26$. The resulting $g^{(1)}$ profiles (see Fig. \ref{fig:5}a-d) are very similar to those obtained at higher $P/P_{th}$. 
At the lowest excitonic fraction explored in our experiments, ${|X|}^2$ = 0.38, a single coherence oscillation is observed, followed by the QSS regime (Fig.~\ref{fig:5}a). 
With increasing ${|X|}^2$, the coherence oscillations double and become more sharply defined (see Fig.~\ref{fig:5}b-d), similar to the case of increasing $P/P_{th}$ in Fig.~\ref{fig:3}. Increasing ${|X|}^2$ (and consequently $P_{th}$) results in the appearance of multiple, well defined coherence oscillations in the TCO regime.

The fact that similar features in the TCO regime can be obtained by tuning different parameters (pump power and excitonic fraction), both of which affect polariton interaction and energy relaxation rates, suggests that these effects originate from early-stage dynamics of the coherence formation, which cannot be directly probed using time-averaging measurements. Consequently, we perform numerical modeling of the experiment to investigate the early-stage dynamics of the coherence development. 

\section{Numerical Modeling}

\begin{figure*}[t!]
    \centering
    \includegraphics[width=\textwidth]{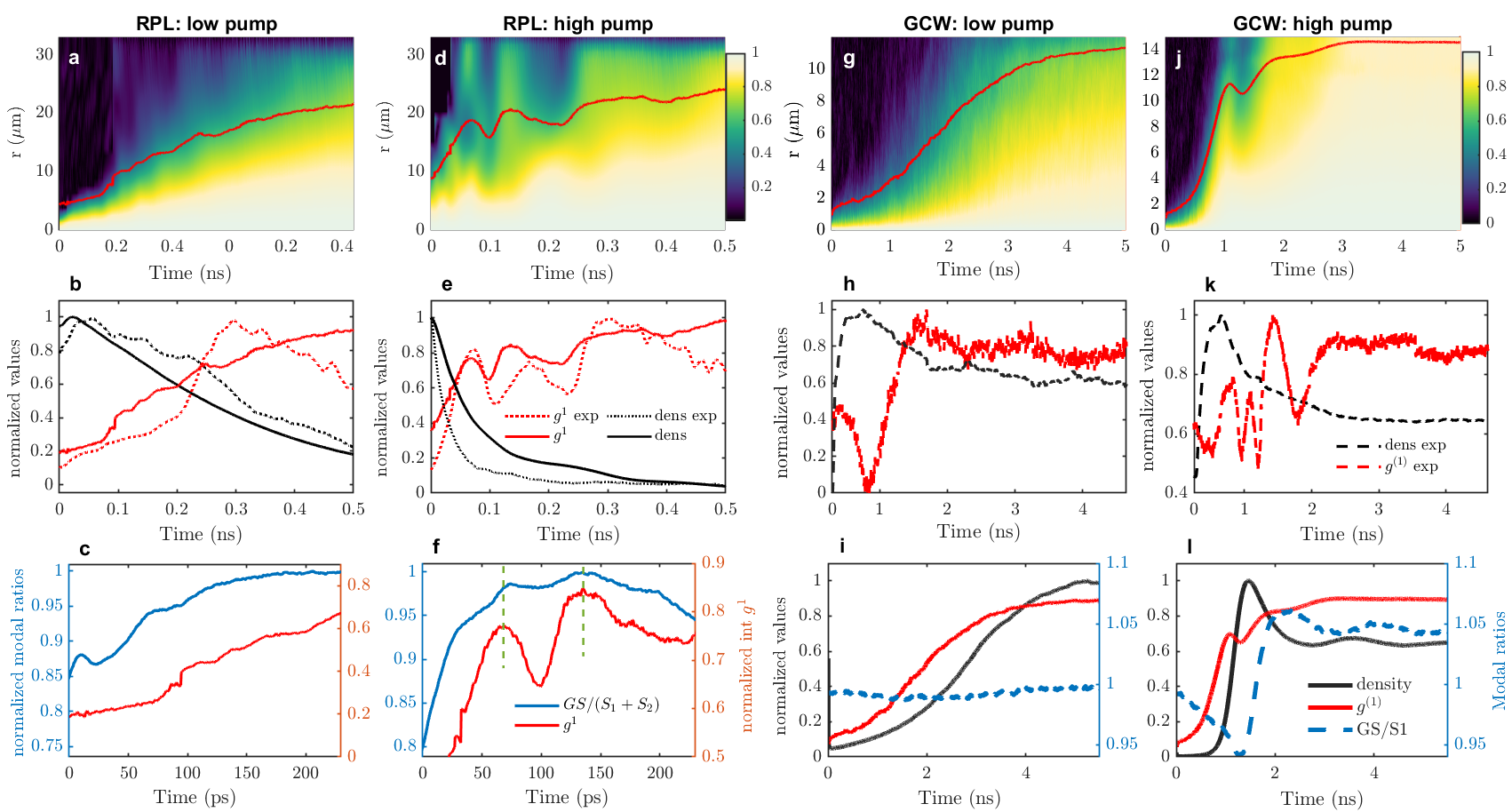}
    \caption{Results of numerical simulations. (a-f) Numerical simulation vs the RPL experiment. (a,d) Numerically simulated time-resolved $g^{(1)}$ profile heatmap and its integrated value (red curve) for the a) low and d) high $P/P_{th}$ ratio. (b,e) Temporal evolution of the integrated density of the condensate (black) and integrated $g^{(1)}$ (red) for the b) low and e) high $P/P_{th}$ ratio. Dashed and solid lines show experimental data and numerical simulations,
respectively. (c,f) Numerically modeled temporal evolution of the modal ratio between the ground (GS) and first excited states ($S_1$, $S_2$) of the condensate (black line) and integrated $g^{(1)}$ (red line) for the c)
low and f) high $P/P_{th}$ ratio. Green dashed lines in (f) indicate the peaks in the modal ratio, which correspond to the oscillations in $g^{(1)}$. (g-l) Numerical simulation vs the GCW experiment. (g,j) Numerically simulated time-resolved $g^{(1)}$ profile heatmap and its integrated value (red line) for the g) low and j) high $B_{p}/B_{p,0}$ ratio. (h,k) Temporal evolution of the experimentally measured integrated density of the condensate (black) and $g^{(1)}$ (red) for the h) low and k) high $B_{p}/B_{p,0}$ ratio. The origin of the time axis coincides with the arrival of the condensate-breaking pulse. (i,l) Numerically calculated evolution of the integrated condensate density (light blue), integrated $g^{(1)}$ (orange), and the modal ratio between the ground (GS) and the first excited state ($S_1$) for the i) low and l) high $B_{p}/B_{p,0}$ ratio.}
    \label{fig:6}
\end{figure*}
\subsection{The model}

To gain insight into the early-stage dynamics of the polariton condensate and coherence formation, we numerically solve the time-dependent equations for polariton mean-field wavefunction, $\psi(\textbf{r},t)$, coupled to the rate equation for the excitonic reservoir density, $n_{R}(\textbf{r},t)$ by employing the truncated Wigner (TW) approximation, which takes into account quantum fluctuations on top of the mean-field~\cite{carusotto2013quantum}:
\begin{align}
 i \hbar d \psi  &= 
 \begin{aligned}[t] 
  dt\bigg[ 
\left( i \beta - 1 \right) \frac{\hbar^2  \nabla^2}{2 m } 
+ g_c|{\psi} |_{{\mathcal{W}}}^2 +g_R n_R + 
\\
 + \frac{i \hbar }{2} \left( R n_R  - \gamma_c \right) 
\bigg] \psi  + i \hbar dW_c 
\label{eq:GPE_pol} 
\end{aligned}
\\ 
\frac{d}{dt} n_R  &= 
\begin{aligned}[t]
P (\textbf{{r}}) - \left( \gamma_R + R |{\psi} |_{{\mathcal{W}}}^2 \right) n_R 
\label{eq:GPE_res}
\end{aligned}
\end{align}
where $\gamma_c$ and $\gamma_R$ define the decay rates of the polaritons and the excitonic reservoir respectively, and $m$ is the polariton effective mass. To simplify the simulations, we restrict our numerical geometry to one spatial dimension ($x$) for both experimentally relevant configurations.
In \eqref{eq:GPE_pol}, quantum fluctuations on top of the mean-field are accounted for by the space- and time-correlated Wiener noise $ \left< dW_c(x,t)  dW_c(x^\prime,t) ~\right> =0$, $\left< dW_c(x,t)  dW_c^*(x^\prime,t)  \right> = (\gamma_\mathrm{c} + R n_\mathrm{R}({x}))/2 \  \delta_{x,x^\prime}dt$.
The renormalized density $|{\psi}|^2_{\mathcal{W}} \equiv
\left|{\psi} \right|^2 - {1}/{a} $  includes the subtraction of the Wigner commutator contribution, where $a$ is the lattice spacing of the numerical grid~\cite{comaron2018dynamical}.
Parameter $\beta$ sets the rate of energy relaxation~\cite{estrecho2018single,Comaron2024}.
The constants $g_c$ and $g_R$ are the strengths of polariton-polariton and polariton-reservoir interactions respectively; they can be estimated as $g_c = g_{ex}|X|^4$, $g_R = g_{ex}|X|^2$, where $g_{ex}$ is the exciton-exciton interaction strength, and $|X|$ is the excitonic Hopfield coefficient~\cite{deng2010exciton, estrecho2019direct}. 
Parameter $R = R_0 g_c/g_R$ quantifies the stimulated scattering rate of the reservoir excitons into the polariton condensate. 
We simulate the experiment using parameters consistent with those of the experimental setup~\cite{Comaron2024}, reported in the Supplemental Document \cite{SM}. The configurations of the two numerical experiments differ in their pumping conditions, which in turn determine the effective confinement.

In RPL configuration, the condensate is formed in a ring-shaped potential and its dynamics is triggered by a pulsed excitation. 

The pulsed ring-shaped excitation is simulated setting an initial condition for the excitonic reservoir
\begin{equation}
n_\mathrm R(x,t=0) =
n_\mathrm R^0 
\left[ e^{-\frac{\left(\lvert x \rvert - r\right)^2}{2\sigma^2}}
+
\bar{n}\right]
\end{equation}
where $r$ and $\sigma$, are the radius and radial width of the ring respectively, $n_\mathrm R^0$ the initial reservoir density, effectively proportional to the initial pulse power $\mathcal{P}^\mathrm{RPL}$ such that $n_\mathrm{R}^0 = \mathcal{P}^\mathrm{RPL}n_\mathrm{R}^\mathrm{th}$, with $n_\mathrm{R}^\mathrm{th}$ the initial reservoir for the polariton threshold, and $\bar{n} = n_{\min} \, \Theta\!\left(r - \lvert x \rvert\right)$ accounts for a low but non-zero density of the reservoir in the centre of the trap~\cite{Pieczarka2019,Comaron2024}. 
To avoid spurious flows of polaritons outside the trap, we also implement a hard--wall potential $V_{\mathrm{trap}} = 0$ if $|x| < r$.

The initial condition is a quasi-vacuum, i.e. a random complex field $\psi(\mathbf{r})$. This Gaussian-distributed random field reproduces the Wigner representation of the 'empty' photonic/excitonic modes, providing the spontaneous scattering seeds that later get amplified by gain into a macroscopic condensate. The system is subsequently evolved according to Eqs.~(\ref{eq:GPE_pol}) and (\ref{eq:GPE_res}). Following the excitation pulse, polaritons undergo decay while flowing toward the center of the ring, exhibiting nontrivial relaxation dynamics. To characterize this behaviour, we monitor the evolution of the density $|\psi|^2$ and the first-order correlation function $g^{(1)}$.

In contrast, GCW configuration is realized under continuous pumping with a Gaussian spatial profile, leading to a driven-dissipative steady-state regime. 
The revival of the condensate after its depletion due to the Stark effect is modeled, for simplicity, via a sudden quench from a fully depleted condensate. This approach captures the physical scenario observed experimentally, as illustrated in Fig.~\ref{fig:4}, and reflects the fact that in both RPL and GCW configurations the condensate grows from an initial quasi-vacuum state. Note that the variation of the CW pump power relative to the condensation threshold in numerical simulations effectively corresponds to varying the strength of the quench in the GCW experiment because a more powerful condensate-breaking pulse in the experiment has an effect of enhancing exciton reservoir density, as discussed above.

The continuous-wave (CW) experiment is modeled using a spatially dependent, time-independent (CW) pump of the form
\begin{equation}
    P = \mathcal{P}^{\mathrm{GCW}} \exp\left(-\frac{x^2}{2\sigma^2}\right),
\end{equation}
where $\mathcal{P}^{\mathrm{GCW}}$ denotes the pump strength and $\sigma$ represents the width of the Gaussian profile. 
The simulations are performed using the same parameter set as the experimental setup considered in this work, consistent with Ref.~\cite{panico2023}, as detailed in Supplemental Document \cite{SM}.

\subsection{Effects of the quench strength}

In analogy with the experiments conducted in the RPL configuration, we investigate two distinct pumping regimes corresponding to the low and high excitation powers relative  to the condensation threshold. 
Figs.~\ref{fig:6}a,d display the spatiotemporal evolution of the first-order correlation function in the RPL configuration. The red lines, corresponding to the spatially-integrated $g^{(1)}$, highlight the emergence of oscillatory behaviour in the high-power regime. 
The time evolution of the spatially-integrated and normalized  density $|\psi|^2$ and correlation function $g^{(1)}$ is shown in Figs.~\ref{fig:6}b,e, demonstrating good qualitative agreement with the experimental observations.

The highly non-equilibrium, driven-dissipative nature of polariton condensation means that the energy relaxation towards the ground state of the condensate is preceded by the regime where multiple energy states of polaritons (confined modes of the effective trap) are simultaneously macroscopically populated.  By transforming to energy--momentum space, we resolve the polariton dispersion, extract the modal composition, and, through energy filtering, isolate individual modes to analyze the temporal evolution of their populations. This analysis is presented in Fig.~\ref{fig:6}c,f for the low- and high-pump regimes, respectively.
In the low-pump case, Fig.~\ref{fig:6}c, the modal populations exhibit a predominantly monotonic growth, with only weak oscillations at low coherence. 
In contrast, the high-pump regime, Fig.~\ref{fig:6}f, is characterized by high coherence with pronounced oscillations in $g^{(1)}$ and in the lowest modes population.
The oscillatory behavior of the modal population ratios, occurring in phase with the coherence oscillations, suggests that the latter originate from a nontrivial redistribution of polariton population among the modes following the rapid quench across the condensation transition.

We now focus on the simulations reproducing the GCW configuration.
For simplicity, the simulations model the condensate-breaking pulse as an instantaneous reset of the condensate wavefunction to vacuum. The effect of the varying power of the condensate-breaking pulse on the quench is modeled by the varying power of CW excitation instead. This is justified by the fact that the AC Stark effect, while acting resonantly and transiently perturbing the excitonic reservoir, operates on timescales much faster than those of the condensate dynamics, with the pump-injected reservoir density setting the effective timescale of the condensate recovery.
Figs.~\ref{fig:6}g,j display the spatiotemporal evolution of the first-order correlation function for two distinct pumping regimes, corresponding to the CW pump powers above and high above threshold, respectively. 
Consistent with the experimental observations and similarly to the RPL configuration, the spatial coherence exhibits oscillations in the high-power regime.

Further insight into the temporal dynamics is provided by the evolution of the spatially-integrated $g^{(1)}$ and $|\psi|^2$, shown in Figs.~\ref{fig:6}i,l and compared with the corresponding experimental measurements for the low and high powers of the condensate-breaking pulse (Figs.~\ref{fig:6}h,k).
As in the previous numerical experiment, we proceed by extracting the populations of the different energy modes. The ratio between the two lowest-energy modes, shown as a blue dashed line, reveals no redistribution of population among lowest-energy modes in the low-pump regime. In contrast, the high-pump regime exhibits a pronounced and dynamic redistribution.

Inspection of the polariton distributions (see Supplemental Document \cite{SM}) indicates the presence of a single mode hopping between two energy states during the system's relaxation dynamics, taking place at about $1.2$~ns. 
We note that while mode hopping near threshold has been reported previously~\cite{alnatah2024critical}, its occurrence during quench dynamics towards a state deep in the ordered phase has not, to our knowledge, been previously documented.
Importantly, in Fig.~\ref{fig:6}l, we observe that the modal redistribution of populations at later times coincides with the oscillations in the integrated coherence. This correlation confirms that the origin of the observed oscillations lies in the interaction-driven population exchange between different energy modes.

\section{Discussion \& Conclusion}

To confirm the origin of the observed TCO regime in the coherence dynamics following a quench, it is important to rule out other mechanisms that may be responsible for coherence oscillations.

The transient oscillatory behavior of the spatial coherence after the quench may also occur due to density modulations in the condensate.
It has been theoretically predicted that oscillatory dynamics can arise in coupled reservoir--polariton systems as a consequence of dynamical instabilities, even in spatially homogeneous configurations~\cite{opala2018}. 
In such cases, these oscillations typically occur in the low-pump regime, close to threshold, where the interplay between gain and loss leads to unstable population dynamics \red{and can occur at a much faster timescales}. 
This behavior is in contrast with our observations, where pronounced oscillations in $g^{(1)}$ emerge at high pump powers, indicating that a different physical mechanism is at play. 

Additionally, breathing-like and sloshing modes following a quench could, in principle, contribute to oscillatory behavior. 
However, these effects can be ruled out in our system (see Supplemental Document \cite{SM}). In the RPL configuration, it is known that low-energy oscillations can occur under certain conditions~\cite{Estrecho2021low}. Therefore, in our experiment we deliberately limited the excitonic fraction range to avoid persistent collective oscillations of the condensate density~\cite{Estrecho2021low}. When present, such density oscillations feature many cycles lasting more than 1~ns, which is in stark contrast to the single or few-cycle coherence oscillations observed in this work (see Supplemental Document \cite{SM}). Furthermore, even when there are density oscillations, the careful analysis of the $g^{(1)}$ shows that the coherence oscillation is decoupled from the density oscillation (see Fig. S10 in the Supplementary Document \cite{SM}). 

Finally, by further exploring the numerical model (see Supplemental Document Section 3 \cite{SM}), we find that the oscillations increase both in number and amplitude with increasing energy relaxation parameter $\beta$. 
This trend highlights the crucial role of relaxation processes in enhancing the dynamical redistribution of population among modes, reinforcing our interpretation of the mechanism underpinning coherence oscillations. 

The two-stage dynamics of coherence development, with the TCO and QSS regimes,  is consistently observed in two different experiments, independently performed by two research groups at different locations. 
In the RPL configuration, a trapped condensate is created by a pulsed optical excitation and is allowed to decay without continuous replenishment. 
In GCW configuration, an expanding steady-state condensate is created via CW pumping and a laser pulse was introduced to momentarily extinguish this condensate. 

Both configurations revealed oscillations in $g^{(1)}$ at early times. This transient regime is strongly influenced by the pumping power, $P/P_{th}$, the breaking pulse power, and the excitonic fraction. 
Numerical simulations allowed us to attribute the coherence oscillations to the redistribution of population between the excited and ground state which randomizes the relative phase within the condensate resulting in a decrease in coherence. 
The increased interparticle interactions, reservoir-condensate interactions and faster scattering processes which are consequences of high pumping or a more excitonic detuning, make the early-time condensation dynamics chaotic as reflected by the multiple population redistribution events corresponding to clearer and more frequent $g^{(1)}$ oscillations.
This transient phase is then followed by the build-up of a spatially uniform $g^{(1)}$. This coherence is maintained under the CW excitation or decays together with the condensate in the pulsed excitation regime.

In summary, the emergence of spatial coherence in polariton condensates follows a structured dynamical route starting with a transient oscillation phase which is largely interaction-driven, followed by a stable coherence phase. The consistency of this behavior across two very different experimental configurations highlights the importance of the early time condensation dynamics in shaping spatial coherence within the condensate. 

More broadly, our results demonstrate that the emergence of long-range coherence in a driven-dissipative condensate can proceed through a distinct transient phase in which coherence is repeatedly established and destroyed before a stable ordered state is reached. The observed oscillations reveal that phase ordering is intimately linked to the dynamical redistribution of polariton population among competing condensate modes, a process that remains hidden in conventional density measurements that average over multiple condensate realizations. 
This finding suggests that the route to coherence following a quench cannot always be described as a simple monotonic growth of correlations, but may instead involve transient mode competition and repeated reorganization of the condensate wavefunction.
Similar mechanisms are expected to arise in other non-equilibrium bosonic systems where interactions, relaxation, and gain-loss processes coexist, including photon condensates~\cite{klaers2010photons} and driven atomic Bose gases ~\cite{Billam_2012, Diehl}. The ability to directly resolve these dynamical processes through time-dependent spatial coherence measurements opens a new avenue for investigating the microscopic pathways through which quenched quantum fluids establish macroscopic order and approach universal coherent states.

\begin{backmatter}
\bmsection{Funding} Australian Research Council (CE170100039, DE220100712). 

National Science Centre (Poland)/Narodowe Centrum Nauki 2020/39/D/ST3/03546. 

The Princeton portion of this work is funded in part by the Gordon and Betty Moore Foundation EPiQS initiative, Grant GBMF14260.

\bmsection{Acknowledgments} This work was financially supported within the "NP Research, innovation and competitiveness for green and digital transition 2021-2027" (PN RIC 2021-2027) co-financed by the European Union, the Italian Ministry of Enterprises and Made in Italy (MIMIT) and the Italian Ministry of University and Research (MUR), through projects: AI-PHOQUS "Artificial Intelligence and Advanced Networks Embedded in Photonics and Quantum Sciences and Technology", CUP B83C26000470007; "Centro meridionale per l'innovazione quantistica", CUP: B49H26000300007; and "Polo meridionale di innovazione per l'informazione quantistica", CUP: B42F26000460005. The authors acknowledge funding also from the Italian Ministry of University and Research (MUR) under the granting scheme FIS 3 (grant number FIS-2024-04047) and from the European Innovation Council (EIC) Pathfinder Programme in the frame of the European Union’s Horizon Europe initiative, through projects: "Quantum Optical Networks based on Exciton-polaritons" (Q-ONE), Horizon-EIC-2022-Pathfinder Challenges, grant agreement No. 101115575 and "Neuromorphic Polariton Accelerator" (PolArt), Horizon-EIC-2023-Pathfinder Open, grant agreement No. 101130304. Views and opinions expressed were however those of the authors only and do not necessarily reflect those of the European Union or EIC and SMEs Executive Agency (EISMEA). Neither the European Union nor the granting authority can be held responsible for them.

\bmsection{Supplemental document}
See Supplement 1 for supporting content. 

\end{backmatter}

\bibliography{reference}

\end{document}


\maketitle

\section{Details of experiment under RPL configuration: Time-resolved spatial coherence of optically-trapped exciton-polariton condensates}

This experiment was done in the Polariton Bose-Einstein Condensate (PolBEC) laboratory in the Australian National University, Canberra ACT 2601, Australia.

\subsection{Sample Information, Detuning, \& Excitation Power}

The sample used in this work is an ultrahigh-quality $3\lambda/2$ GaAs-based microcavity, consisting of distributed Bragg reflectors (DBRs) with 32 (top) and 40 (bottom) pairs of alternating Al$_{0.2}$Ga$_{0.8}$As/AlAs layers and an active region of 12 GaAs/AlAs quantum wells of 7 nm nominal thickness ~\cite{steger2015slow}. It is held in a continuous flow cryostat (Janis ST-500) with a temperature maintained at $<$10~K.

\begin{figure}[htbp]
\centering
\includegraphics[width = 0.85\linewidth]{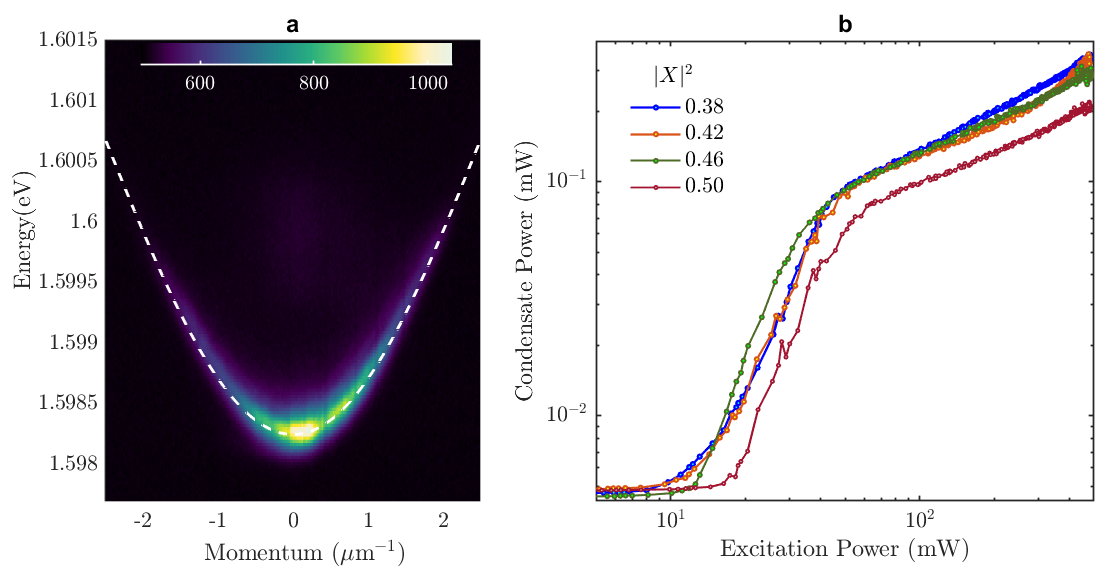}
\caption{a) Momentum-resolved spectrum of polariton emission measured at a very low excitation power to extract the dispersion (energy-momentum) relation (dashed line) of polaritons. b) Log-log plot of the input-output power relation of the four sample detunings probed in this work.}
\label{fig:Dispersion}
\end{figure}

The microcavity in our sample has a wedge that translates to the linear gradient of the cavity photon energy, and therefore the exciton-photon detuning, $\delta$, across the sample. The detuning $\delta$ is defined as the difference between the cavity photon mode ($E_{cav}$) and the exciton resonance ($E_{exc}$) as shown in the equation below,

\begin{equation}
    \delta = E_{cav} - E_{exc}.
\end{equation}

It controls the excitonic and photonic composition of the polariton condensate through the Hopfield coefficients, $X$ and $C$, as follows,  \cite{DengCavLifetime}

\begin{equation}
    |X|^2 = \frac{1}{2} \left(1 + \dfrac{\delta}{\sqrt{\delta^2 + \Omega^2}}\right), \quad and \quad |C|^2 = 1 - |X|^2
\end{equation}
where $|X|^2$ and $|C|^2$ are the excitonic and photonic fractions respectively, and $\Omega$ is the Rabi splitting. In other words, the detuning $\delta$ controls how photon-like or exciton-like the polariton condensate is and consequently controls the effective mass of polaritons in the lower polariton branch as follows,

\begin{equation}
    \frac{1}{m_{LP}} = \frac{|X|^2}{m_{exc}} + \frac{|C|^2}{m_{cav}},
\end{equation}
where $m_{LP}$ is the lower polariton's effective mass, and $m_{exc}$ and $m_{cav}$ are the exciton mass and cavity photon effective mass, respectively \cite{DengCavLifetime}.
Near zero and at slightly positive detuning, the increasing $|X|^2$
 makes lower polaritons heavier \cite{DengCavLifetime} and enhances interparticle interactions \cite{Comaron2024,estrecho2019direct, sun2017direct}, resulting in more efficient scattering processes \cite{estrecho2019direct}. This facilitates rapid relaxation into the ground state, leading to a more efficient population buildup and reduced condensation threshold, $P_{th}$. Pushing toward very positive detunings, however, results to a higher critical density for polariton condensation, stronger exciton localization, and shorter coherence time for the lower polaritons. All of these hinder efficient polariton condensation \cite{DengCavLifetime}. Conversely, at negative detuning where $|C|^2$ dominates, $m_{LP}$ is lighter and the critical density for polariton condensation is reduced. However, with more photon-like characteristics, the scattering processes are less efficient causing the polaritons to accumulate at higher momentum states \cite{Tassone1997, Coles2013}, and oftentimes these polaritons decay without reaching the ground state. To overcome this relaxation bottleneck, higher pumping powers are often required \cite{Tartakovskii2000}. In summary, the detuning is an important parameter as it directly influences the interparticle interactions and the rate and efficiency of the scattering and relaxation processes necessary for polariton condensation in the ground state \cite{DengCavLifetime, estrecho2019direct, Comaron2024}. 

In this work, we investigated the influence of detuning on the establishment of spatial coherence within the condensate. We probed four regions of the sample with detunings of -3.9 meV ($|X|^{2}$~=~0.38), -2.5 meV ($|X|^{2}$~=~0.0.42), -1.1 meV ($|X|^{2}$~=~0.46), and 0 ($|X|^{2}$~=~0.50). The cavity photon lifetime, $\tau_{C}$, is estimated to range from 206-230~ps \cite{Steger_polaritonlifetime} in the order of the least to the most excitonic detuning. Fig.~\ref{fig:Dispersion}a shows a typical energy-momentum relation for 0 detuning at a very low excitation power. The white-dashed lines trace the extracted lower polariton dispersion, which is the basis for estimating the sample detuning.

Fig.~\ref{fig:Dispersion}b shows the input-output power relations of the four sample detunings. The condensate's emission intensity non-linearly increases as a function of excitation power, signifying polariton condensation. We define the condensation threshold, $P_{th}$, as the excitation power corresponding to the onset of this non-linear increase in emission intensity. The rest of the excitation powers used in this work are defined in terms of their ratio to $P_{th}$. Notice that a slight variation in the excitonic fraction,$|X|^{2}$, results in a noticeable change in the input-output power relation. This means that each detuning has a different effective excitation range even if the same excitation power is used. 

\subsection{Experimental Set-ups}
\subsubsection{Creation of the dynamical polariton condensate}

To facilitate polariton condensation in an optically induced trap, a femtosecond pulse-laser (Coherent Chameleon Ultra II), chopped by an acousto-optic modulator (AOM) at 1$\%$ duty cycle, is shaped into a ring of radius 16~$ \mu m$ using an axicon of conical angle 1$^\circ$. This configuration effectively confines the condensate at the center of the optical trap, resulting to the minimal overlap between condensate and excitonic reservoir \cite{askitopoulos2013polariton,estrecho2018single,Askitopoulos2019}. The annular excitation beam is then directed to the sample in the cryostat via reflection from the dichroic mirror and focused by a 0.5 NA objective. The sample is pumped off-resonantly with the excitation beam tuned to the lowest reflectance minimum of the microcavity (719 nm or 1.72 eV). Note that the pump photon energy is more than 100 meV above the lower polariton energy at ~1.6 eV (~775 nm), ensuring that any coherence from the pump is lost through a complex relaxation process \cite{byrnes2014exciton}. Owing to the pulsed nature of the excitation, the polariton condensate is not continuously replenished with polaritons, and instead evolves and decays within the duration of each pulse. This results in a sharply defined, transient population of polariton condensates, with their dynamics encoded in the emitted light.

\subsubsection{Time-resolved characterization of the condensate emission}

\begin{figure}[htbp]
\centering
\includegraphics[width = 1.0\linewidth]{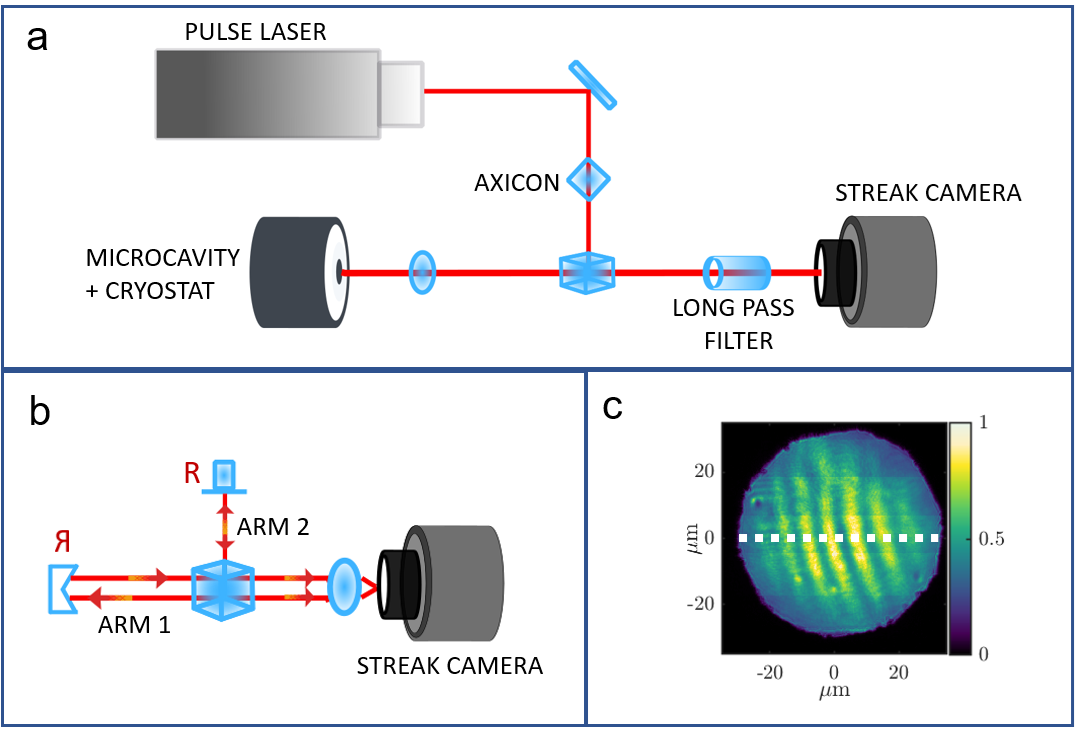}
\caption{a) Schematics of the experimental set-up for measuring the temporal evolution of the polariton condensate. b) Schematics of the Michelson interferometry set-up coupled to a streak camera to measure the temporal evolution of spatial coherence. (c) Real space image of interference fringes. The white-dashed lines indicate the region that is imaged by the streak camera.}
\label{fig:ExpSetup}
\end{figure}

The polariton condensate's emission is collected by the same objective and is transmitted through the same dichroic mirror mentioned in the previous section. The collected light passes through a long-pass filter to block any residual laser contribution and then is directed to the streak camera (Optronis OptoScope SC-10) to capture the time evolution of the condensate's intensity. For measuring the condensate's energy redshift upon decay, the collected and filtered light is directed to a spectrometer coupled to a streak camera. Fig.~\ref{fig:ExpSetup}a shows the schematics for these measurements.

\subsubsection{Time-resolved $1^{st}$ order spatial coherence measurement}

To measure the temporal evolution of spatial coherence of the condensate, the Michelson interferometry set-up is coupled to the streak camera. As shown in Fig. \ref{fig:ExpSetup}b, a $50/50$ beam splitter splits the condensate emission into two equal beams (arm 1 and arm 2) with one undergoing retro-reflection (arm 1). One of the mirrors is mounted on a piezoelectric translation stage which enables the minute variations in the path length between these two beams and produces a relative phase between them. These two arms are then overlapped to form vertical fringes and then sent to the streak camera for time-resolved imaging. Fig. \ref{fig:ExpSetup}c shows a real-space image of the interference pattern created. The white dashed-line indicates the portion that goes through the slit of the streak camera. 
In this configuration, the time delay between two beams is kept to almost zero and therefore the $1^{st}$ order correlation function, $g^{(1)}$, 

\begin{equation}
g^{(1)}(\Delta x, t,\Delta t =0) = \frac{\langle \Psi(x,t)   \Psi^\dagger(-x,t) \rangle}{\sqrt{\langle |\Psi(x,t)|^2 \rangle \langle  \Psi^\dagger(-x,t)|^2 \rangle}},
\end{equation}

represents the spatial coherence as a function of time $t$ relative to the arrival of the laser pulse. Here, $\Psi$ represents the condensate wavefunction and $\Delta x$ is the distance between two points in the condensate measured from the center.
Given some interference intensity pattern, the visibility, $V$, is given by

\begin{equation}
V = \frac{I_{max} - I_{min}}{I_{max}+I_{min}},
\end{equation}

where $I_{max}$ and $I_{min}$ corresponds to the maximum and minimum intensity of the fringe pattern, respectively. At the detector, the intensity, $I$, is given by

\begin{equation}
I = I_{1} + I_{2} + 2 \sqrt{I_{1}I_{2}}|g^{(1)}| cos\phi,
\end{equation}

where $I_{1}$ and $I_{2}$ are the intensities of arms 1 and 2, respectively and $\phi$ is the relative phase. Following this notation,

\begin{equation}
I_{max} = I_{1} + I_{2} + 2 \sqrt{I_{1}I_{2}}|g^{(1)}| cos\phi
\end{equation} 
and
\begin{equation}
I_{min} = I_{1} + I_{2} - 2 \sqrt{I_{1}I_{2}}|g^{(1)}| cos\phi,
\end{equation}

and therefore,

\begin{equation}
V = \frac{2\sqrt{I_{1}I_{2}}}{I_{1}+I_{2}} |g^{(1)}|.
\end{equation}

In our case, the two beam arms are identical ($I_{1} = I_{2}$), which means that
\begin{equation}
    V = |g^{(1)}|.
\end{equation}

We extracted the visibility values of the interference pattern to obtain the time-resolved $g^{(1)}$ profiles.

\subsection{Temporal evolution of spatial coherence at various detunings}

As mentioned earlier, four different sample detunings where investigated in this work. For each of these detunings, the temporal evolution of $g^{(1)}$ were measured at increasing pump power, $P/P_{th}$, and the results are shown below.

Fig.~\ref{fig:38Excitonic} shows the time-resolved $g^{(1)}$ profiles for a sample detuning of $|X|^{2}~=~0.38$. This is the most photonic detuning probed in this work. The TCO regime appears starting at P~=~4.50~$P_{th}$. As $P/P_{th}$ increases, the QSS regime is established faster as indicated by the decreasing time interval between the two coherence regimes.

\begin{figure}[htbp]
\centering
\includegraphics[width = 1.0\linewidth]{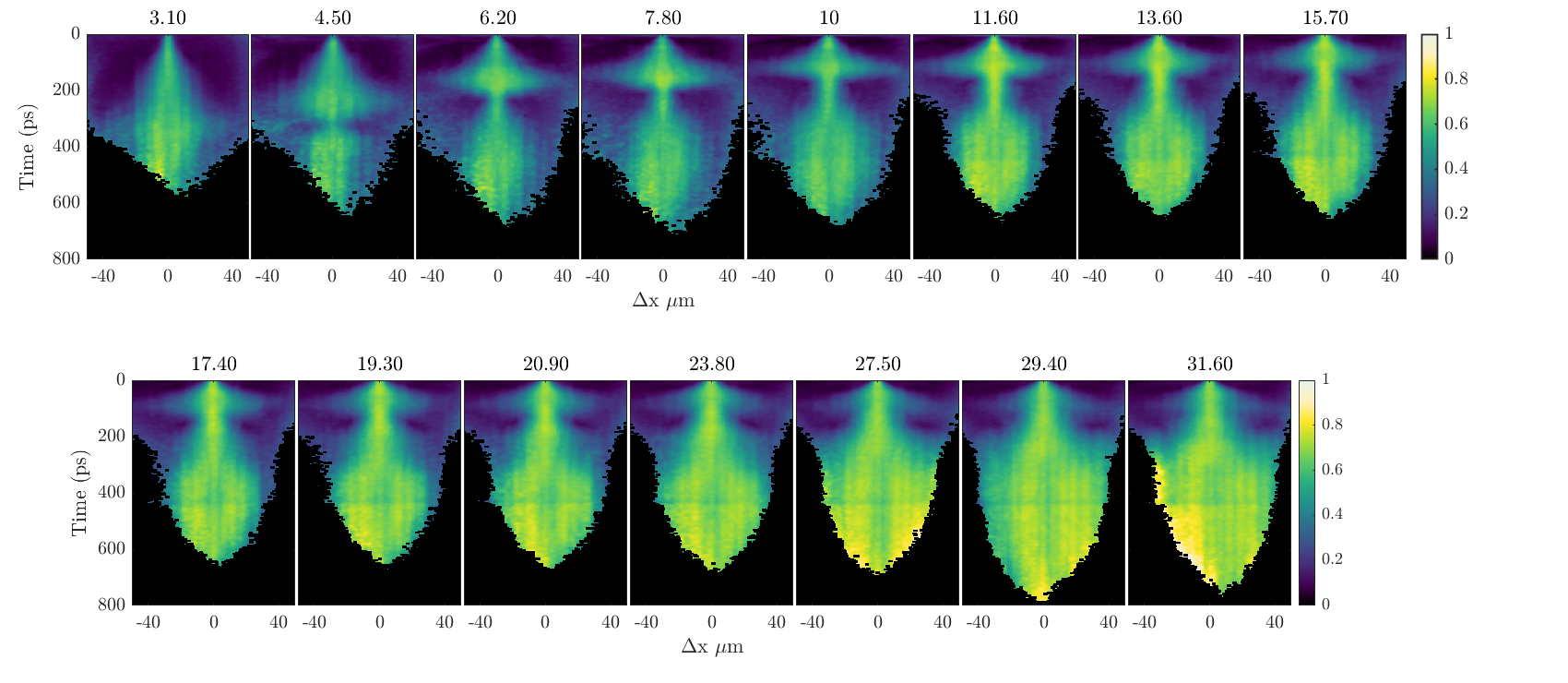}
\caption{Time-resolved $g^{(1)}$ profiles for a sample detuning of $|X|^2$~=~0.38 at increasing $P/P_{th}$.}
\label{fig:38Excitonic}
\end{figure}

\begin{figure}[htbp]
\centering
\includegraphics[width = 1.0\linewidth]{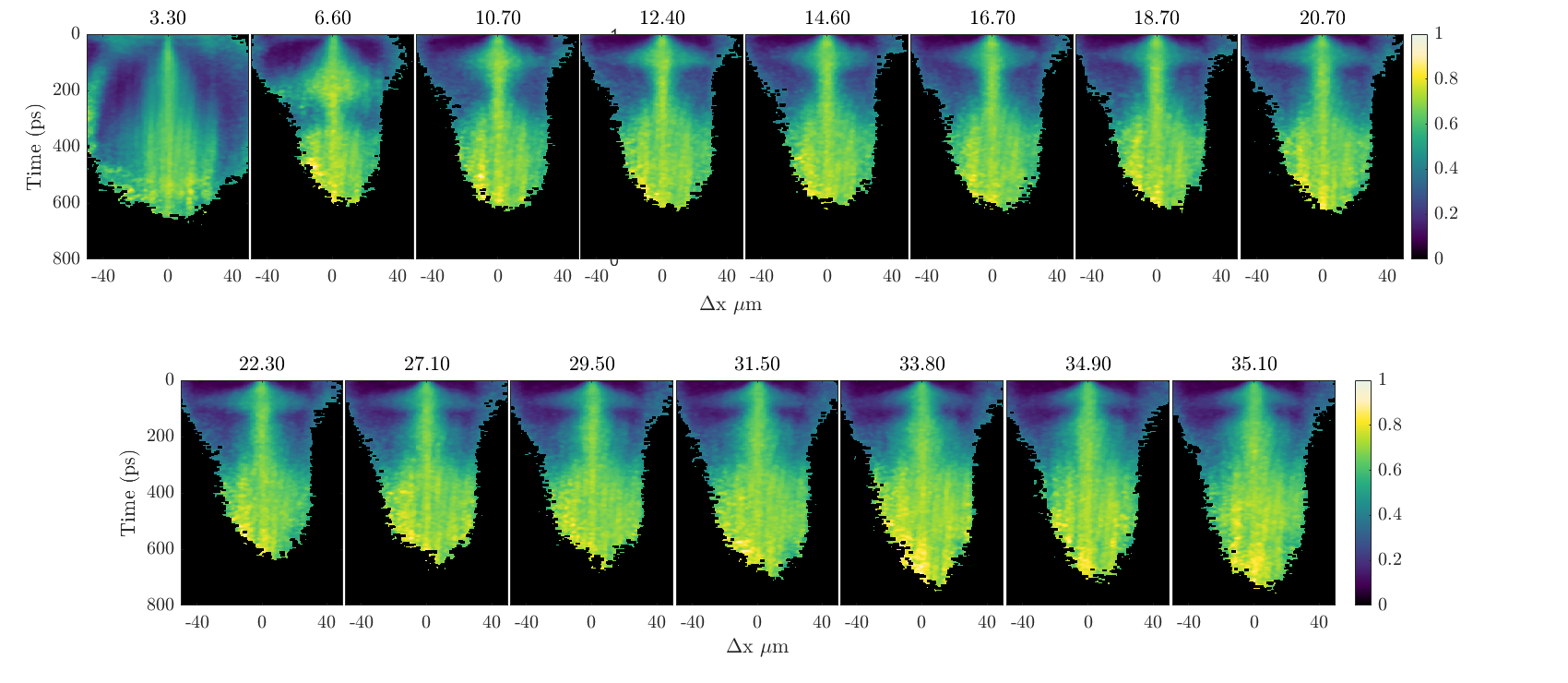}
\caption{Time-resolved $g^{(1)}$ profiles for a sample detuning of $|X|^2$~=~0.42 at increasing $P/P_{th}$.}
\label{fig:42Excitonic}
\end{figure}

Fig.~\ref{fig:42Excitonic} shows the time-resolved $g^{(1)}$ profiles for a sample detuning of $|X|^{2}~=~0.42$. The profiles show similar characteristics with the sample detuning $|X|^{2}~=~0.38$, wherein only one coherence oscillation can be clearly seen in the TCO regime. Although the time interval between the TCO and QSS regimes decreases with increasing $P/P_{th}$, coherence gradually improves throughout the QSS regime, with $g^{(1)}$ values being lower at early times and then increases at later times. 

\begin{figure}[htbp]
\centering
\includegraphics[width = 1.0\linewidth]{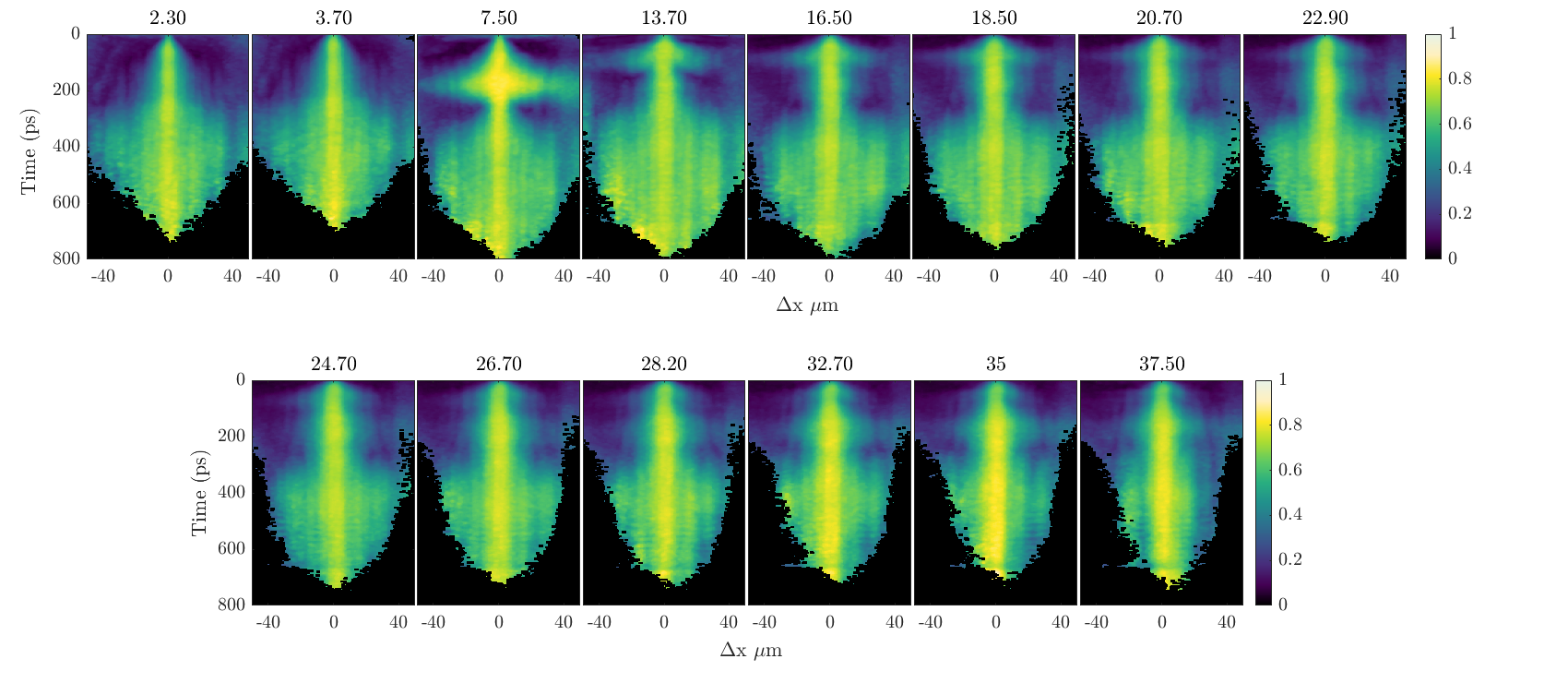}
\caption{Time-resolved $g^{(1)}$ profiles for a sample detuning of $|X|^2$~=~0.46 at increasing $P/P_{th}$.}
\label{fig:46Excitonic}
\end{figure}

Fig.~\ref{fig:46Excitonic} shows the time-resolved $g^{(1)}$ profiles for a sample detuning of $|X|^{2}~=~0.46$. Here the coherence profiles exhibit very high $g^{(1)}$ values at the center even in the QSS regime. With increasing $P/P_{th}$, there is an increase in coherence oscillations in the TCO regime and the coherence length is significantly shorter during this time. The QSS regime is also not spatially uniform with the central region being more coherent (brighter, visually) than at the sides.

\begin{figure}[htbp]
\centering
\includegraphics[width = 1.0\linewidth]{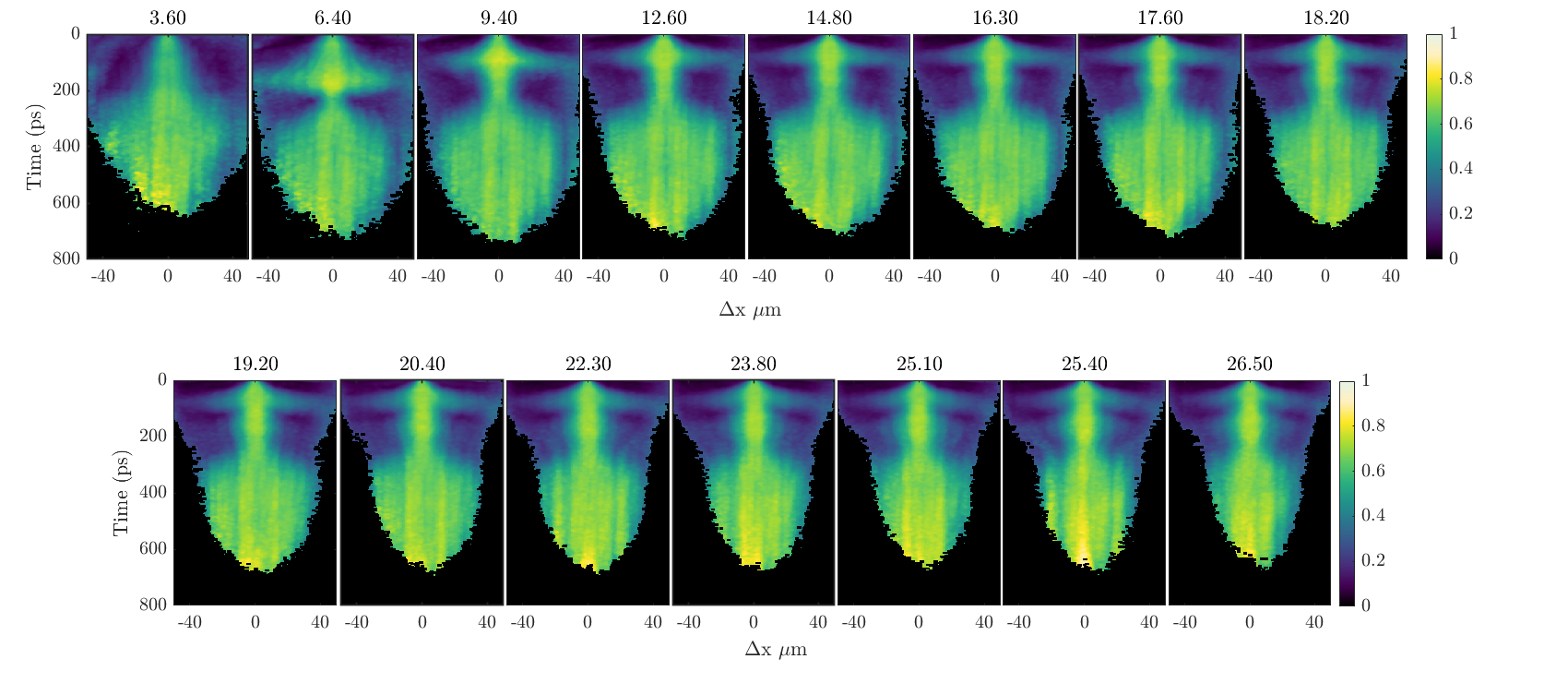}
\caption{Time-resolved $g^{(1)}$ profiles for a sample detuning of $|X|^2$~=~0.50 at increasing $P/P_{th}$.}
\label{fig:50Excitonic}
\end{figure}

Fig.~\ref{fig:50Excitonic} shows the time-resolved $g^{(1)}$ profiles at increasing pump power, $P$, for a sample detuning of $|X|^{2}~=~0.50$. This is the most excitonic detuning probed in our work. Here, multiple coherence oscillations in the TCO regime can be clearly seen. Beyond P~=~9~$P_{th}$, the time interval between the two coherence regimes remains approximately the same.

In all detunings sampled, both the TCO and QSS regimes are observed. Furthermore, the TCO regime does not appear if $P/P_{th}$~<~4. It is also clear that increasing the pumping power makes the TCO regime more turbulent as seen in Figs. \ref{fig:42Excitonic}, \ref{fig:46Excitonic}, and \ref{fig:50Excitonic}. These similarities greatly outweigh the minute differences that can be seen in the $g^{(1)}$ profiles of the four sample detunings, which strongly suggest that the same mechanism is responsible for these coherence effects.

\begin{figure}[htbp]
\centering
\includegraphics[width = 1.0\linewidth]{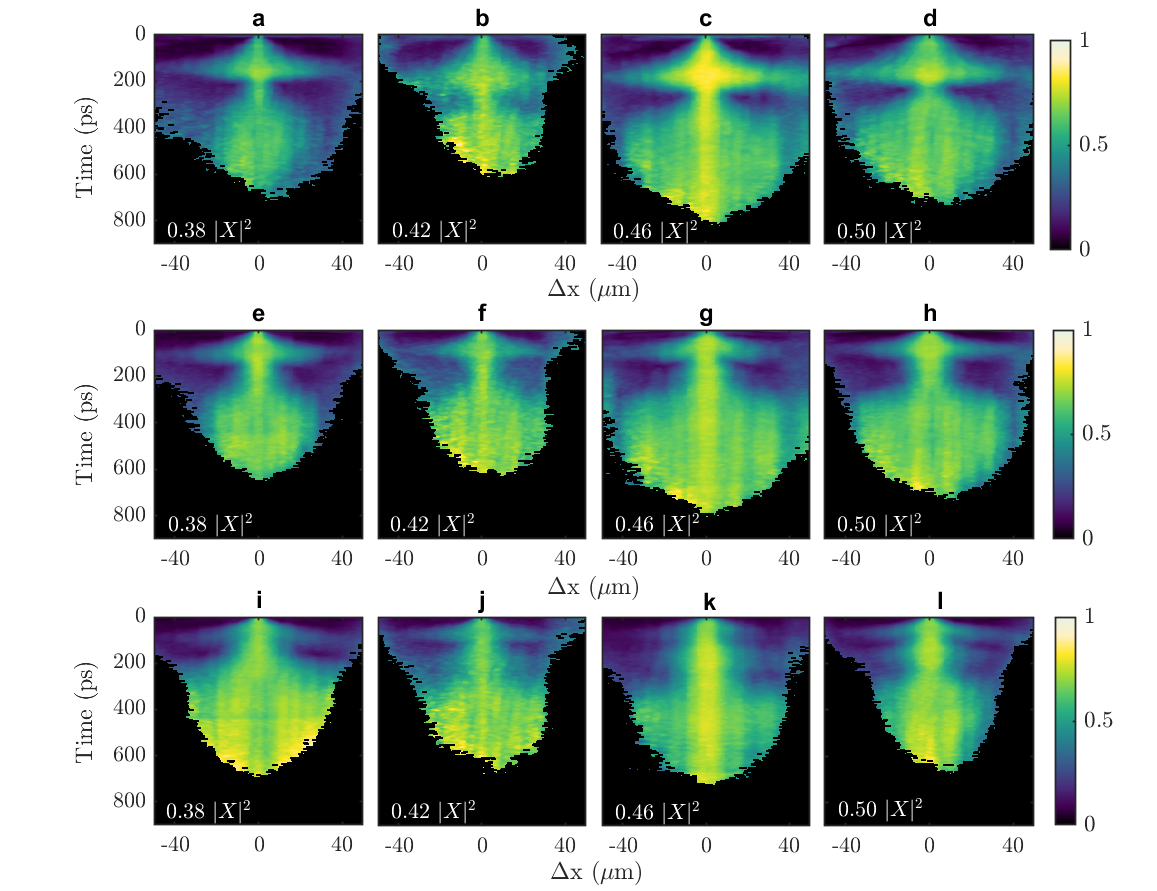}
\caption{Time-resolved $g^{(1)}$ profiles at pumping power (a-d) 7~$P/P_{th}$, (e-h) 13~$P/P_{th}$, and (i-l) 27~$P/P_{th}$.}
\label{fig:all}
\end{figure}

Note that a slight change in detuning shifts the polariton condensation threshold, $P_{th}$ \cite{PhysRevLett.101.146404, DengCavLifetime}. Therefore to make the comparison more accurate, the experiment was conducted in such a way that all four sample detunings were optically excited around the same power with respect to its unique threshold, $P/P_{th}$. The complete results are shown in Fig. \ref{fig:all}. Here, we observed that the combination of increasing $|X|^2$ and $P/P_{th}$ leads to the increase of coherence oscillations in the TCO regime.

\subsection{Temporal evolution of the coherence length, $\ell_{\rm coh}$}

The coherence length, $\ell_{\rm coh}$, is related to $g^{(1)}$ as shown below,

\begin{equation}
    g^{(1)}(\Delta x,t) \sim e^{-\frac{\Delta x}{\ell_{\rm coh}}}
\label{eq:cohlength_supp}
\end{equation}

\begin{figure*}[htbp]
\centering
\includegraphics[width = 1.0\textwidth]{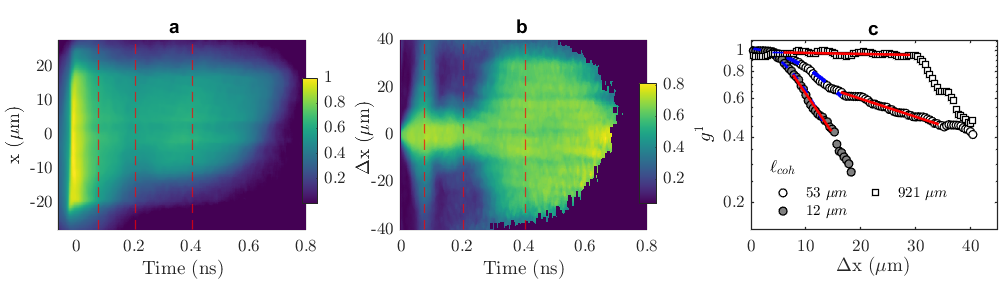}
\caption{Time-resolved a) condensate PL and (b) $g^{(1)}$ for a sample detuning of $|X|^2~=~0.50$ at $\sim$~16~$P_{th}$. The red-dashed lines mark different spatial coherence characteristics in time. Spatial $g^{(1)}$ slices taken from the center of the profile (maximum visibility) towards the periphery at the times specified by the red lines in Fig.~\ref{fig:cohlength}a,b. The blue dashed-lines are the gaussian fits, while the red solid lines are the exponential fits for coherence length, $\ell_{\rm coh}$, estimation.}
\label{fig:cohlength}
\end{figure*}

\begin{figure*}[htbp]
\centering
\includegraphics[width = 1.0\textwidth]{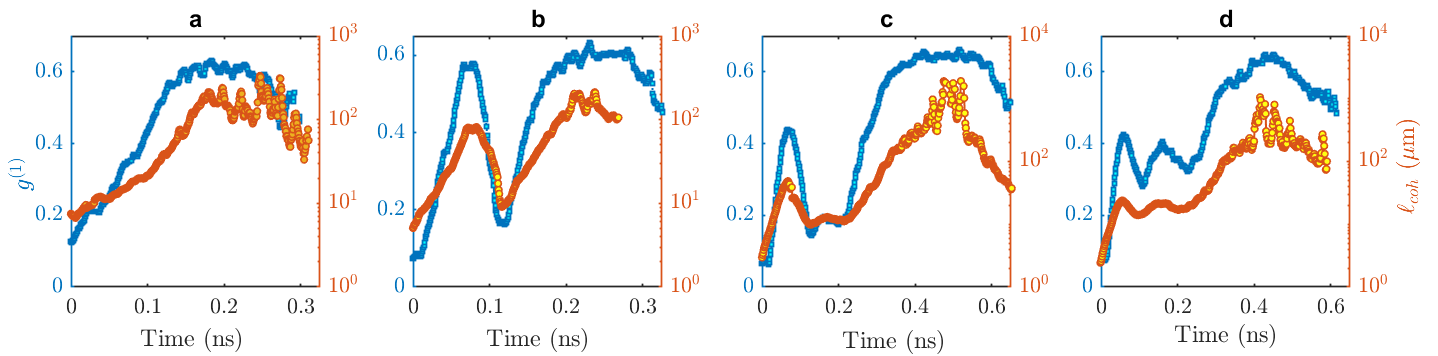}
\caption{Temporal evolution of $g^{(1)}$ (blue) and coherence length, $\ell_{\rm coh}$, (orange) for polariton condensate of sample detuning $|X|^2$~=~0.50 pumped at a) 4, b) 6, c) 16, and d) 26~$P/P_{th}$ in RPL configuration.}
\label{fig:cohlength_powerscan}
\end{figure*}

In this work, $\ell_{\rm coh}$ is extracted by fitting Eq. (\ref{eq:cohlength_supp}) to constant-time slices of $g^{(1)}$, as illustrated in Fig. \ref{fig:cohlength}c. We observe pronounced variations in $\ell_{\rm coh}$ across distinct spatial coherence regimes. To illustrate this, three representative times were selected (red dashed lines in Fig. \ref{fig:cohlength}a,b) and the corresponding coherence times, $\ell_{\rm coh}$, were estimated. As shown in Fig. \ref{fig:cohlength}c, in the TCO regime and at the peak of the first coherence burst ($t$~=~78~ps), the coherence length is $\ell_{\rm coh}$~=~53~$\mu$m, comparable to the condensate size. Immediately after the collapse of this burst, $\ell_{\rm coh}$ decreases to 12~$\mu$m, approximately four times smaller than the condensate. In contrast, in the QSS regime at $t$~=~504~ps, the coherence length increases to $\ell_{\rm coh}$~=~921~$\mu$m, exceeding the condensate diameter by more than an order of magnitude. As shown in Fig. \ref{fig:cohlength_powerscan}, the temporal evolution of the coherence length closely follows that of $g^{(1)}$, wherein the same trends and effects are clearly seen with increasing pump power.

 \subsection{Temporal evolution of spatial coherence of an oscillating condensate}
\label{subsec:oscillatingcondensate}

\begin{figure}[htbp]
\centering
\includegraphics[width = 1.0\linewidth]{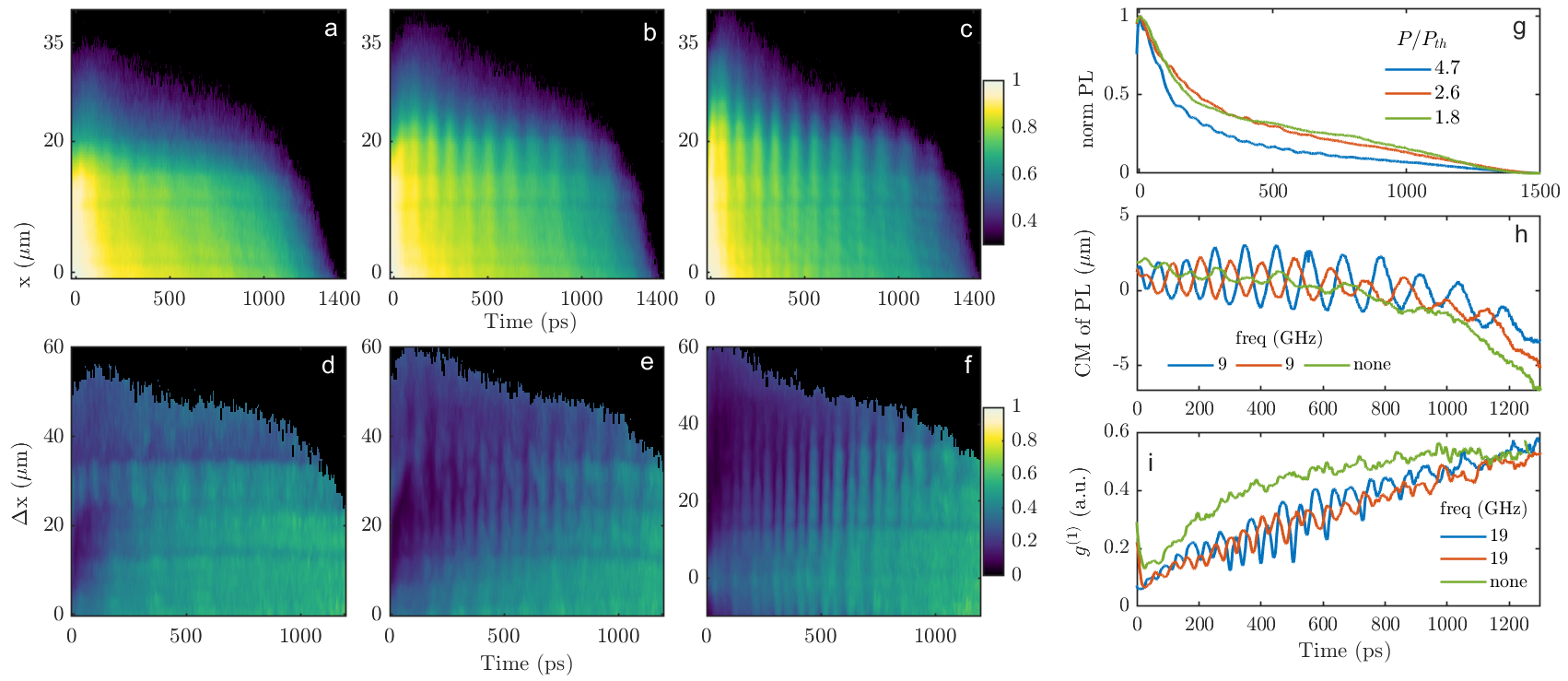}
\caption{Dynamics of a sloshing condensate. Time-resolved condensate PL and $g^{(1)}$ for a pump power of (a \& d) 4.7~$P/P_{th}$, (b \& e) 2.6~$P/P_{th}$, and (c \& f) 1.8~$P/P_{th}$. g) Temporal evolution of the condensate PL. h) Oscillations in the condensate PL with respect to the center of mass (CM). i) Oscillations in $g^{(1)}$.}   
\label{fig:intentionaloscillations}
\end{figure}

It is important to rule out the possibility that these coherence oscillations in the TCO regime resulted from the evaluation of g$^{(1)}$ of a condensate that is spatially sloshing over time. Since the measured time-resolved condensate PL profile is a sum of a thousand realizations, one cannot safely conclude that there are no oscillations in the condensate density as these could be washed out in the measurement. To exclude this possibility, we intentionally manipulated the excitation beam to create a condensate with density oscillations \cite{MilenaDGiorgi2014} that can be clearly captured by the experimental set-up (see Fig.~\ref{fig:intentionaloscillations}c), and then measured the time-resolved $g^{(1)}$ profile (see Fig.~\ref{fig:intentionaloscillations}f).
This case is achieved using an excitation pump with $\lambda$~=~705~nm, pump power of $4.7~P/P_{th}$, and with a sample detuning of $|X|^2~=~0.59 $. 

The dependence of these density oscillations and the corresponding $g^{(1)}$ to pump power is also probed and these are shown in Figs.~\ref{fig:intentionaloscillations}(a-c) and \ref{fig:intentionaloscillations}(d-f), respectively. Clearly, the oscillations in both density and $g^{(1)}$ appear and are enhanced with increasing pump power. For $P/P_{th}~=~1.8$, no oscillations can be clearly seen (Fig.~\ref{fig:intentionaloscillations} a \&  d). Furthermore, the oscillations in $g^{(1)}$ mirror that of the condensate density as shown in Figs.~\ref{fig:intentionaloscillations}i and \ref{fig:intentionaloscillations}h, with an oscillation frequency of 9~GHz and 19~GHz for the density and $g^{(1)}$, respectively. In this case, the dynamics in $g^{(1)}$ can be well attributed to the condensate's spatial distribution over time.

Although these induced density oscillations produce oscillations in g$^{(1)}$, it clearly did not replicate the structured coherence dynamics that reflects both the TCO and QSS regimes. The visibility values (hence,$g^{(1)}$) are generally lower, and the coherence oscillations persist until the condensate is fully decayed. Therefore, a sloshing condensate cannot explain the presence of these two coherence regimes.

\section{FULL TEMPORAL DYNAMICS OF THE CONDENSATE UNDER GCW CONFIGURATION}

This experiment was done in the Advanced Photonics Laboratory in the Institute of Nanotechnology, Lecce, Italy. 

\begin{figure}[htbp]
\centering
\includegraphics[width = 1.0\linewidth]{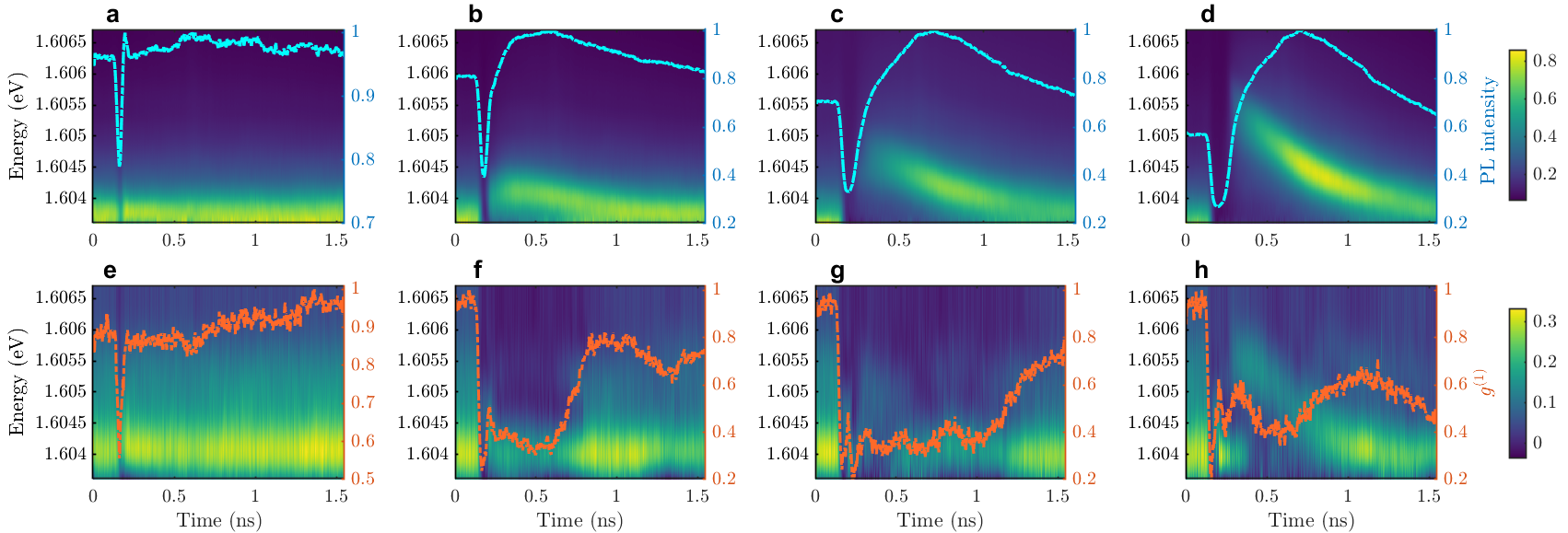}
\caption{Time-resolved condensate PL and $g^{(1)}$ heatmaps for the pump power ratio $B_{p}/B_{p,0}$ of (a,e) 1, (b,f) 2, (c,g) 4, and (d,h) 8 in the TCO regime of the GCW experiment. The blue line is the condensate density evolution quantified by integrating the PL intensity at each time and normalizing it by the peak value. The orange line is the spatial coherence integrated over the vertical axis and normalized to its peak value.}   
\label{fig:TCO_GCW}
\end{figure}

The main text focused on the temporal evolution of the condensate's density (represented by the photoluminescence intensity) and spatial coherence ($g^{(1)}$) after the arrival of the pulse that quenches the condensate under the GCW configuration. In Fig.~\ref{fig:TCO_GCW}, we show the full temporal dynamics, starting from the creation of the steady-state condensate (under CW excitation), in the TCO regime for both the aforementioned quantities.

\section{Numerical details}

\subsection{Spectral decomposition}

To gain further insight into the origin of the observed coherence dynamics, we analyze the time evolution of the dispersion relation. This is obtained by applying a two-dimensional Fourier transform in space and time to the simulated field, yielding the energy- and momentum-resolved distribution $n_{k}(k,E)$. From this quantity, we can resolve the instantaneous dispersion and extract the population of the different energy modes, particularly in regimes where temporal oscillations of the first-order coherence $g^{(1)}$ are observed.

\begin{figure}[htbp]
\centering
\includegraphics[width = 1\linewidth]{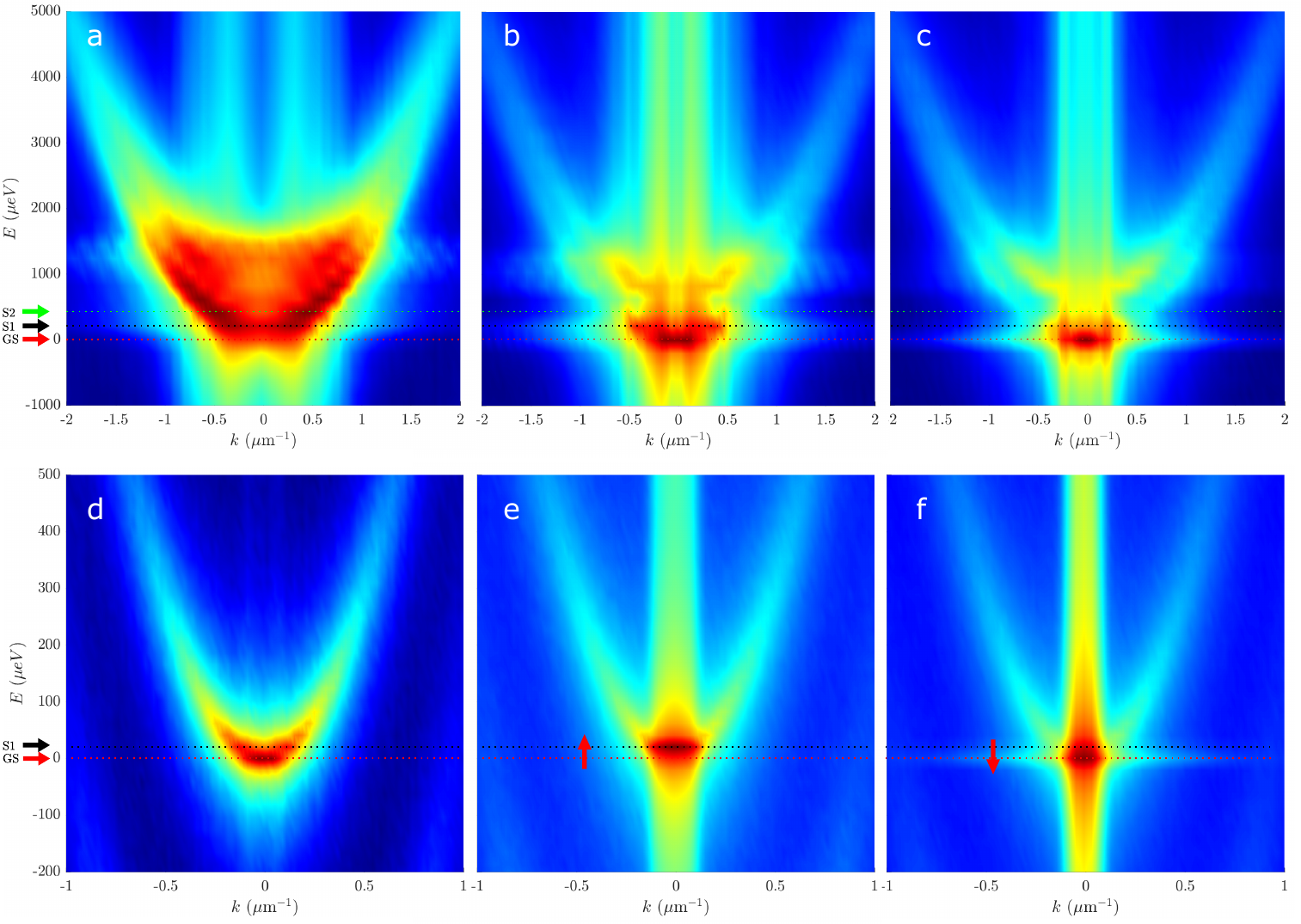}
\caption{\textbf{Spectral decomposition in the numerical simulations}. 
(a-c): Experiment with pulsed excitation on a non-resonant ring. 
For the case $\mathcal{P}^\mathrm{pls}_\mathrm{2}$, the dispersion relation $n_{k}(E,k)$ is depicted for three different times: a) $0\,\mathrm{ps}$, $80\,b) \mathrm{ps}$, and c) $180\,\mathrm{ps}$. 
Colored arrows indicate the energy of the individual modes.
(c-e): Experiment with CW laser. 
For the case $\mathcal{P}^\mathrm{cw}_\mathrm{2}$, the dispersion relation $n_{k}(E,k)$ is depicted for three different times: d) $0\,\mathrm{ps}$, e) $1.2\,\mathrm{ns}$, and f) $2\,\mathrm{ns}$. 
Colored arrows indicate the energy of the individual modes.}
\label{fig:num_spectral_decomp}
\end{figure}

For the pulsed excitation case [Fig.~\ref{fig:num_spectral_decomp} (a-c)], the system initially displays a broad occupation in momentum space, with significant population across a wide range of $k$. A narrow dispersion associated with polaritons excited on top of the ring-shaped pump region is also visible in the spectra. 
As time evolves, the system relaxes towards the minimum of the dispersion while the overall population decays. The distributions $n_{k}(E,k)$ shown in the figure are normalized to highlight the redistribution of population among the modes.
For the CW excitation case, the dispersion are depicted in Fig.~\ref{fig:num_spectral_decomp}(d-f).

By extracting the population of the individual states, we can directly relate the coherence oscillations to the relative occupation of the modes, in both experiments. As discussed in the main text, the analysis suggests that the temporal behavior of $g^{(1)}$ is determined by the ratio between the populations of the dominant states (see Fig.~6 (c,f) for the RPL experiment and Fig.~6 (i,l) for the GCW experiment).

It is interesting to note that, in the CW excitation case, the smaller energy spacing between modes leads to a stronger coupling and more efficient population transfer. As a consequence, after an initial relaxation towards the ground state, the system exhibits pronounced population redistribution among modes, resulting in a dynamical mode-hopping behavior. This effect reflects the enhanced competition between closely spaced energy levels and plays a key role in shaping the observed coherence dynamics.

\subsection{Further numerical tests}

In order to identify the physical origin of the observed oscillations in the first-order coherence, we perform additional numerical tests aimed at ruling out the contribution of breathing-like modes or other density-driven collective excitations.

To this end, we repeat the simulations described in the main text using identical initial conditions, while introducing controlled fluctuations in the pump. Specifically, we consider a noisy pump of the form
\begin{equation}
P = P_0 \left(1 + 0.5\,\xi \right),
\end{equation}
where $\xi$ is a stochastic variable whose seed is correlated with that of the single realization. This procedure mimics small variations of the pump strength within a realistic experimental range (about 5\%).

We observe that, under these conditions, the small-amplitude density oscillations appearing at early times—possibly originating from the repulsive interactions between polaritons and the exciton reservoir (see oscillations in a range $t<50$~ps in Fig.~6k of the main text) —are strongly suppressed. This indicates that such oscillations are sensitive to perturbations and are not a robust feature of the dynamics. 

In contrast, although partially reduced, the oscillations in the first-order coherence $g^{(1)}$ remain clearly visible. The persistence of these coherence oscillations, even in the absence of well-defined density modulations, demonstrates that they are not directly linked to breathing modes or other collective density oscillations.

These results therefore allow us to exclude breathing-like dynamics as the primary mechanism behind the observed oscillations in $g^{(1)}$, supporting instead their interpretation as arising from the interaction-driven redistribution of population among competing spectral modes.

 \newpage
\bibliography{reference}